\documentclass[preprint,1p]{elsarticle}
\usepackage{amsmath,hyperref,mhchem,nicefrac,color,float,lineno,upgreek}

\restylefloat{table}

\newcommand{\Offline}{\mbox{$\overline{\rm Off}$\hspace{.05em}\raisebox{.3ex}{$\underline{\rm line}\enspace$}}}
\DeclareRobustCommand{\OfflineT}{\mbox{$\overline{\rm  Off}$\hspace{.05em}\raisebox{.3ex}{$\underline{\rm  line}\,$}}}

\journal{Astroparticle Physicss}

\begin{document}

\begin{frontmatter}
    \title{The EUSO-SPB2 Fluorescence Telescope for the Detection of Ultra-High Energy Cosmic Rays}
    \author[uah]{James~H.~Adams~Jr.}
\author[apc]{Denis~Allard}
\author[uah]{Phillip~Alldredge}
\author[lehman]{Luis~Anchordoqui}
\author[catania,palermo]{Anna~Anzalone}
\author[apc,torinoINFN]{Matteo~Battisti}
\author[moscow]{Alexander~A.~Belov}
\author[torinoINFN,torinoUniversity]{Mario~Bertaina}
\author[marshall]{Peter~F.~Bertone} 
\author[apc]{Sylvie~Blin-Bondil}
\author[uah]{Julia~Burton}
\author[bariINFN]{Francesco~S.~Cafagna}
\author[infnRome,romeUniversity,riken]{Marco~Casolino}
\author[olomouc]{Karel~\v{C}ern\'{y}}
\author[marshall]{Mark~J.~Christl}
\author[infnNaples,naplesUniversity]{Roberta~Colalillo}
\author[berkley]{Hank~J.~Crawford}
\author[apc]{Alexandre~Creusot}
\author[pennstate]{Austin~Cummings}
\author[UChicago]{Rebecca~Diesing}
\author[infnNaples,naplesUniversity]{Alessandro~Di~Nola}
\author[riken]{Toshikazu~Ebisuzaki}
\author[UChicago]{Johannes~Eser}
\author[torinoINFN,torinoUniversity]{Silvia~Ferrarese}
\author[UChicago]{George Filippatos\corref{cor1}}\ead{gfil@uchicago.edu}
\author[mines]{William~W.~Finch}
\author[infnNaples,naplesUniversity]{Flavia~Flaminio}
\author[rome2]{Claudio~Fornaro}
\author[mines]{Duncan~Fuehne}
\author[kth]{Christer~Fuglesang}
\author[iowa]{Diksha~Garg}
\author[torinoINFN,torinoUniversity]{Alessio~Golzio}
\author[infnNaples,naplesUniversity]{Fausto~Guarino}
\author[apc2]{Claire~Gu\'epin}
\author[mines]{Tobias~Heibges}
\author[berkley]{Eleanor~G.~Judd}
\author[moscow]{Pavel~A.~Klimov}
\author[goddard]{John~F.~Krizmanic}
\author[mines]{Viktoria~Kungel}
\author[iowa]{Luke~Kupari}
\author[uah]{Evgeny~Kuznetsov}
\author[torinoINFN,torinoUniversity]{Massimiliano~Manfrin}
\author[warsaw]{Włodzimierz~Marsza{\l}}
\author[utah]{John~N.~Matthews}
\author[infnNaples,naplesUniversity]{Marco~Mese}
\author[UChicago]{Stephan~S.~Meyer}
\author[torinoINFN]{Marco~Mignone}
\author[torinoINFN,torinoUniversity]{Hiroko~Miyamoto}
\author[moscow]{Alexey~S.~Murashov}
\author[iowa]{Jane~M.~Nachtman}
\author[columbia]{Angela~V.~Olinto}
\author[iowa]{Yasar~Onel}
\author[infnNaples]{Giuseppe~Osteria}
\author[infnNaples,naplesUniversity]{Beatrice~Panico}
\author[apc,cc]{\`Etienne~Parizot}
\author[lehman]{Tom~Paul}
\author[olomouc]{Miroslav~Pech}
\author[infnNaples]{Francesco~Perfetto}
\author[hb]{Lech~W.~Piotrowski}
\author[infnRome,romeUniversity]{Zbigniew~Plebaniak}
\author[iowa]{Jonatan~Posligua}
\author[apc]{Guillaume~Pr\'ev\^ot}
\author[warsaw2]{Marika~Przybylak}
\author[uah]{Patrick~Reardon}
\author[iowa]{Mary~Hall~Reno}
\author[ee]{Marco~Ricci}
\author[mines]{Fred~Sarazin}
\author[bb]{P.~Schov\'{a}nek}
\author[infnNaples,naplesUniversity]{Valentina~Scotti}
\author[warsaw]{Kenji~Shinozaki}
\author[lehman]{Jorge~F.~Soriano}
\author[UChicago]{Ben~K.~Stillwell}
\author[warsaw3]{Jacek~Szabelski}
\author[riken]{Yoshiyuki~Takizawa}
\author[apc,moscow]{Daniil~Trofimov}
\author[kth]{Fredrik~Unel}
\author[infnNaples,naplesUniversity]{Laura~Valore}
\author[goddard]{Tonia~M.~Venters}
\author[uah]{John~Watts~Jr.}
\author[mines]{Lawrence~Wiencke}
\author[mines]{Hannah~Wistrand}
\author[marshall]{Roy~Young}

\affiliation[uah]{
    organization={University of Alabama in Huntsville},
    addressline={301 Sparkman Dr NW}, 
    city={Huntsville},
    state={AL},
    postcode={35899},
    country={United States of America}
}
\affiliation[apc]{
    organization={Université de Paris, CNRS, AstroParticule et Cosmologie},
    addressline={10 Rue Alice Domon et Léonie Duquet},
    city={Paris},
    postcode={75013},
    country={France}
}
\affiliation[lehman]{
    organization={Lehman College City University of New York (CUNY)},
    addressline={250 Bedford Park Boulevard West},
    city={Bronx},
    state={NY},
    postcode={10468},
    country={United States of America}
}
\affiliation[catania]{
    organization={Istituto Nazionale di Fisica Nucleare},
    addressline={Via S. Sofia, 64},
    city={Catania},
    state={CT},
    postcode={95125},
    country={Italy}
}
\affiliation[palermo]{
    organization={Istituto di Astrofisica Spaziale e Fisica Cosmica},
    addressline={Via Ugo la Malfa, 153},
    city={Palermo},
    state={PA},
    postcode={90146},
    country={Italy}
}
 \affiliation[torinoINFN]{
    organization={Istituto Nazionale di Fisica Nucleare},
    addressline={Via Pietro Giuria, 1},
    city={Torino},
    state={TO},
    postcode={10125},
    country={Italy}
}
\affiliation[moscow]{
    organization={Skobeltsyn Institute of Nuclear Physics, Lomonosov Moscow State University},
    addressline={1c2},
    city={Moscow},
    postcode={119991},
    country={Russia}
}
\affiliation[torinoUniversity]{
    organization={Dipartimento di Fisica, Università di Torino},
    addressline={Via P. Giuria 1},
    city={Torino},
    state={TO},
    postcode={10125},
    country={Italy}
}
\affiliation[marshall]{
    organization={NASA Marshall Space Flight Center},
    addressline={Martin Rd SW},
    city={Huntsville},
    state={AL},
    postcode={35808},
    country={United States of America}
}
 \affiliation[bariINFN]{
    organization={Istituto Nazionale di Fisica Nucleare},
    addressline={Via Giovanni Amendola, 173},
    city={Bari},
    state={BA},
    postcode={70126},
    country={Italy}
}
 \affiliation[infnRome]{
    organization={Istituto Nazionale di Fisica Nucleare},
    addressline={Via della Ricerca Scientifica},
    city={Rome},
    state={RM},
    postcode={00133},
    country={Italy}
 }
 \affiliation[romeUniversity]{
    organization={Università di Roma Tor Vergata},
    addressline={Via della Ricerca Scientifica, 1},
    city={Rome},
    state={RM},
    postcode={00133},
    country={Italy}
 }
\affiliation[riken]{
    organization={RIKEN},
    addressline={2-1 Hirosawa},
    city={Wako},
    state={Saitama},
    postcode={351-0198},
    country={Japan}
}
\affiliation[olomouc]{
    organization={Palacký University},
    addressline={Křížkovského 511/8},
    city={Olomouc},
    postcode={779 00},
    country={Czech Republic}
}
 \affiliation[infnNaples]{
    organization={Istituto Nazionale di Fisica Nucleare},
    addressline={Strada Comunale Cinthia},
    city={Napoli},
    state={NA},
    postcode={80126},
    country={Italy}
 }
 \affiliation[naplesUniversity]{
    organization={Università Federico II - Dipartimento di Fisica},
    addressline={Via Cintia, 21 - Building 6},
    city={Napoli},
    state={NA},
    postcode={80126},
    country={Italy}
 }
\affiliation[berkley]{
    organization={Space Science Laboratory, University of California},
    addressline={7 Gauss Way},
    city={Berkeley},
    state={CA},
    postcode={94720},
    country={United States of America}
}
\affiliation[pennstate]{
    organization={The Pennsylvania State University},
    addressline={4575 Pollock Rd},
    city={State College},
    state={PA},
    postcode={16801},
    country={United States of America}
}
\affiliation[UChicago]{
    organization={The University of Chicago},
    addressline={5801 S. Ellis Ave},
    city={Chicago},
    state={IL},
    postcode={60637},
    country={United States of America}
}
\affiliation[mines]{
    organization={Colorado School of Mines},
    addressline={1500 Illinois St},
    city={Golden},
    state={CO},
    postcode={80401},
    country={United States of America}
}
\affiliation[rome2]{
    organization={Università telematica internazionale Uninettuno},
    addressline={Corso Vittorio Emanuele II, 39},
    city={Rome},
    state={RM},
    postcode={00186},
    country={Italy}
}
\affiliation[kth]{
    organization={KTH Royal Institute of Technology},
    addressline={Brinellvägen 8},
    city={Stockholm},
    postcode={114 28},
    country={Sweden}
}
\affiliation[goddard]{    
    organization={NASA Goddard Space Flight Center},
    addressline={8800 Greenbelt Rd},
    city={Greenbelt},
    state={MD},
    postcode={20771},
    country={United States of America}
}
\affiliation[warsaw]{
    organization={National Centre for Nuclear Research},
    addressline={Andrzeja Sołtana 7/3},
    city={Otwock},
    postcode={05-400},
    country={Poland}
}
\affiliation[utah]{
    organization={University of Utah},
    addressline={115 S 1400 E},
    city={Salt Lake City},
    state={UT},
    postcode={84112},
    country={United States of America} 
}
\affiliation[iowa]{
    organization={University of Iowa},
    addressline={30 N Dubuque St},
    city={Iowa City},
    state={IA},
    postcode={52242},
    country={United States of America}
}
\affiliation[apc2]{
    organization={Laboratoire Univers et Particules de Montpellier Universite de Montpellier, CNRS/IN2P3},
    addressline={Pl. Eugene Bataillon},
    city={Montpellier},
    postcode={34090},
    country={France}
}
\affiliation[columbia]{
    organization={Columbia University},
    addressline={535 West 116th St},
    city={New York},
    state={NY},
    postcode={10027},
    country={United States of America}
}
\affiliation[cc]{
    organization={Institut Universitaire de France (IUF)},
    addressline={Saint-Michel, 103},
    city={Paris},
    postcode={75005},
    country={France}
}
\affiliation[hb]{
    organization={Faculty of Physics, University of Warsaw},
    addressline={ Ludwika Pasteura 5},
    city={Warsaw},
    postcode={02-093},
    country={Poland}
}
\affiliation[warsaw2]{
    organization={University of Łódź Doctoral School of Exact and Natural Sciences},
    addressline={21/23 Jana Matejki Street},
    city={Łódź},
    postcode={90-237},
    country={Poland}
}
\affiliation[ee]{
    organization={Istituto Nazionale di Fisica Nucleare},
    addressline={Via Enrico Fermi, 54},
    city={Frascati},
    state={RM},
    postcode={00044},
    country={Italy}
}
\affiliation[bb]{
    organization={Institute of Physics of the Czech Academy of Sciences},
    addressline={Na Slovance 1999/2},
    city={Praha},
    postcode={182 00},
    country={Czech Republic}
}
\affiliation[warsaw3]{
    organization={Stefan Batory Academy of Applied Sciences},
    addressline={Stefana Batorego 64C},
    city={Skierniewice},
    postcode={96-100},
    country={Poland}
}

\cortext[cor1]{Corresponding author}
    \begin{abstract}
    The Extreme Universe Space Observatory on a Super Pressure Balloon 2 (EUSO-SPB2) flew on May 13$^{\text{th}}$ and 14$^{\text{th}}$ of 2023. Consisting of two novel optical telescopes, the payload utilized next-generation instrumentation for the observations of extensive air showers from near space. One instrument, the fluorescence telescope (FT) searched for Ultra-High Energy Cosmic Rays (UHECRs) by recording the atmosphere below the balloon in the near-UV with a 1~$\mu$s time resolution using 108 multi-anode photomultiplier tubes with a total of 6,912 channels. Validated by pre-flight measurements during a field campaign, the energy threshold was estimated around 2~EeV with an expected event rate of approximately 1 event per 10 hours of observation. Based on the limited time afloat, the expected number of UHECR observations throughout the flight is between 0 and 2. Consistent with this expectation, no UHECR candidate events have been found. The majority of events appear to be detector artifacts that were not rejected properly due to a shortened commissioning phase. Despite the earlier-than-expected termination of the flight, data were recorded which provide insights into the detectors stability in the near-space environment as well as the diffuse ultraviolet emissivity of the atmosphere, both of which are impactful to future experiments.
\end{abstract}
    \begin{keyword}
        UHECR \sep cosmic ray \sep super pressure balloon \sep fluorescence \sep ultraviolet
    \end{keyword}
\end{frontmatter}

\section{Introduction}

Despite observational advancements, the origin of the highest energy cosmic rays has not yet been identified. The Ultra High Energy Cosmic Ray (UHECR) energy spectrum falls $\propto E^{-2.7}$ above $10^{18.2}$ eV, and even faster at higher energies \citep{Aab_2020}. As a result the highest energy UHECRs require very large collecting areas in order to be detected. Current state-of-the-art UHECR detectors achieve acceptances (apertures) of order $10^4$~km$^2$ sr \citep{auger_exposure} by deploying detectors spread over thousands of square kilometers. In order to detect UHECRs with an aperture that is larger than this by an order of magnitude or more, a new detection strategy is needed \citep{Coleman_2023}. One possibility is to use low cost detectors which can be manufactured at scale, such as radio antennae \citep{grand}. Another is to observe a large volume of the atmosphere from above using a fluorescence detector flown in a low Earth orbit \citep{POEMMA}. In addition to the increase in exposure achieved by observing the atmosphere from above, observations of the longitudinal profile via fluorescence enable estimations of the primary mass of UHECRs \citep{Settimo_2016}. This allows for more precise anisotropy studies \citep{refId0}. While UHECRs are deflected in galactic and extra-galactic magnetic fields, it is expected that particles of high enough rigidity may experience small enough deflections to still carry information about their sources.  

Over the past decade, the JEM-EUSO Collaboration \citep{JEM-EUSO} has pursued several missions working towards the ultimate goal of observing UHECRs from space. The first, EUSO-Balloon \citep{EUSO-Balloon}, flew in 2014 for one night supported by a stratospheric balloon. EUSO-Balloon prototyped a multi anode photomultiplier tube (MAPMT) detector from near space. Given the limited flight duration and aperture, and the lack of an internal trigger, EUSO-Balloon was not expected to make an observation of an UHECR. It did, however, record several hundred speed-of-light tracks in the ultraviolet (UV) initiated by a laser fired from a helicopter flying underneath the balloon. This served as a proof of concept for the observation technique \citep{Abdellaoui_2018}. In 2017, EUSO-SPB aimed to make the first observation of UHECRs via fluorescence from above. Due to a leak developed shortly after reaching float altitude, the mission terminated after 12 days 4 hours afloat before any UHECR could be observed \citep{ABDELLAOUI2024102891}. In addition to the balloon borne experiments, Mini-EUSO \citep{Mini-EUSO} is currently onboard the International Space Station. Looking down on the Earth through a UV-transparent window, Mini-EUSO has made measurements of meteors, transient luminous events and the diffuse UV emissivity of the atmosphere \citep{mini-euso-meteors}. The ground based telescope EUSO-TA \citep{ABDELLAOUI201898} serves as a testbed for the technology used in EUSO experiments and has recorded UHECRs with the help of an external trigger. 

The Extreme Universe Space Observatory on a Super Pressure Balloon 2 (EUSO-SPB2) built on the experience of previous missions undertaken within the framework of the JEM-EUSO Collaboration. EUSO-SPB2 prototyped instrumentation for the hybrid focal surface of the Probe of Extreme Multimessenger Astrophysics (POEMMA) by housing two optical telescopes. The Cherenkov telescope (CT) was designed for the observation of Cherenkov signals close to the shower axis of PeV scale extensive air showers (EAS) \citep{Gazda:2023cj}. A silicon photomultiplier based camera sampled a wide wavelength band at nanosecond timescales. With the ability to tilt above and below the limb, the CT was able to detect EAS initiated by cosmic rays as well as search for EAS created by $\tau$ lepton decays induced by $\nu_\tau$ interactions within the Earth \citep{PhysRevD.104.063029,PhysRevD.103.043017}. The second telescope, the Fluorescence Telescope (FT), utilized MAPMTs in a manner similar to previous JEM-EUSO experiments. Significant advances over previous JEM-EUSO experiments included the use of reflective instead of refractive optics, leading to a higher optical throughput and a more confined point-spread-function, as well as the use of multiple photo detection modules (PDMs) requiring external synchronization and increasing the complexity of the detector.

Due to the nature of the optical observations, EUSO-SPB2 required dark skies for observations. Combined with the solar power required, this necessitates a mid-latitude flight where day night cycles will occur unlike during a polar flight. To maintain a constant altitude through the extreme temperature fluctuations that come with these day night cycles, the balloon must maintain a constant pressure. This is achieved by over-inflating the balloon during launch and sealing both ends. The super pressure balloons utilized by NASA are 18.8 million ft$^3$ (532,000~m$^3$) at float altitude and are designed to support 5,500 lbs (2,500~kg) at 110,000~ft (33.5~km)\footnote{The mission parameters provided by NASA are in imperial units.} altitude. The chosen launch location for this type of mission is W\={a}naka New Zealand, located at -44.7$^\circ$ latitude. By launching during a semi-annual stratospheric wind circulation pattern \citep{stratosphere_wind}, the payload may circle the globe at a relatively constant latitude. The geography of the southern hemisphere means that the majority of this type of trajectory will be over ocean, but there can still be opportunities to terminate the flight over land and recover the payload. NASA has launched five flights of this nature since 2015, with three payloads landing on land allowing for recovery. The second most recent flight, Super-BIT \citep{romualdez2018overview} completed five revolutions before terminating over Argentina on their sixth crossing where the payload landed on solid ground. The decision to terminate the mission was made to guarantee recovery of data, as the predicted trajectory had the balloon drifting south where it may not have crossed land again. However, the balloon could have possibly supported a flight longer than the 39 days achieved. Therefore, Suuper-BIT served as a proof of concept for this method of ultra-long duration mid-latitude balloon flights.
\section{Detector Description}\label{secs:desrciption}

\subsection{Focal Surface}

The focal surface of the FT consisted of three PDMs. The PDM is the base component of the POEMMA focal surface. These PDMs each consisted of nine elementary cells (ECs), each of which was made up of four MAPMTs (Hamamatsu Photonics R11265-M64), for a total of 2,304 pixels per PDM arranged in a square matrix. Each of the nine ECs shared a common high voltage power supply (HVPS) based on a Cockroft-Walton circuit. The ECs were compact assemblies each containing one HVPS generator board and four application-specific integrated circuits (ASICs) for signal digitization. This assembly was potted in a epoxy compound to prevent discharge between the various components. The EC assemblies were roughly 55~mm~$\times$~55~mm. The MAPMTs operated in dual photon counting and charge integration mode (see \cite{KIICRCPaper} for more details), with a gate time unit (GTU) of 1.05~$\mu$s. Each MAPMT was made up of 64 2.88$\times$2.88~mm$^2$ pixels arranged in an 8$\times$8 grid. The overall efficiency of the MAPMTs was roughly 33\% in the wavelength range of interest (300-400~nm). The total field of view (FoV) of the instrument was 3$\times$12$^{\circ}\times12^{\circ}$, with 2$^{\circ}$ separation between each PDM. Projecting this onto the ground at a float altitude of 33~km results in an area of roughly 36~km$^2$. Each individual pixel had a FoV of $\sim$0.23$^{\circ}$. There were gaps between PDMs, ECs and MAPMTs which result in parts of the atmosphere which cannot be observed at a given time.

\begin{figure*}[h!]
    \begin{center}
	    \includegraphics[width=\textwidth]{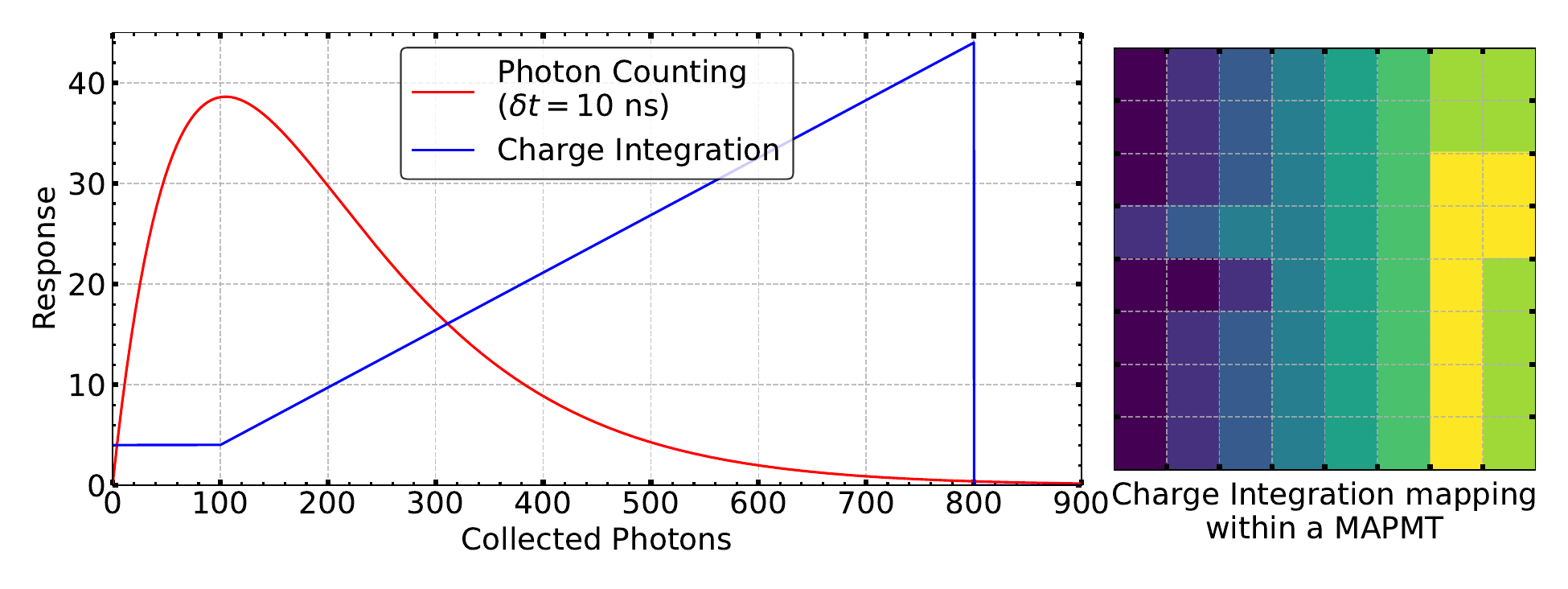}
	    \caption{Response of one pixel of the MAPMT to light at different intensities in both photon counting and charge integration (left) and mapping of charge integration channels within an MAPMT (right). The x-axis ``collected photons'' represents photons which have created a photo-electron from an interaction in the photocathode and that photo-electron has reached the first dynode triggering a particle cascade which can be measured at the anode. The red curve is an analytic function which depends only on the ratio of the integration time ($\tau$) and the double-pulse-resolution ($\delta t$) of the form $N(n)=n\times e^{-n\nicefrac{\delta t}{\tau}}$. The blue curve shows an idealized performance of the charge integration, with the width of the plateau region near 0 photons and the slope of the line being configurable by parameters passed to the ASIC. The irregular groupings of pixels within an MAPMT is the result of routing constraints in the ASIC.}
	    \label{figs:ki_pc}
    \end{center}
\end{figure*}

\subsection{Data Acquisition System}

The EUSO-SPB2 data acquisition system (referred to as DAQ) was derived from the Mini-EUSO \citep{Mini-EUSO_DAQ} and EUSO-SPB1 \citep{SCOTTI201994} systems and worked on a PDM-basis.  Each MAPMT was read out by a SPACIROC3 ASIC \citep{SPACIROC3} that performs both photon counting and charge integration digitization. Photon counting was the main mode of the FT. For each of the 64 pixels in a MAPMT the data is digitized in acquisition windows of 1 GTU, with a double pulse resolution of $\sim$~10~ns. The charge integration is performed integrating the signal from clusters of 8 pixels. The layout of the charge integration channels is shown in right panel of Figure \ref{figs:ki_pc}. The data
 can be used for the research of impulsive bright events (such as the direct Cherenkov signal from showers initiated by $\tau^{\pm}$ decays induced by astrophysical $\nu_{\tau}$).
  
The output of the 36 ASICs of a PDM was multiplexed by three cross boards, each of which contains one Artix 7 FPGA. The three cross boards are connected to the main electronic board, referred to as a Zynq board, containing a Xilinx Zynq 7000 FPGA with an embedded dual core ARM9 CPU processing system. There was one Zynq board for each PDM. The Zynq board controlled the data flow from the ASICs, ran the trigger logics (for the photon counting and charge integrated data), managed the ASIC configurations and interfaced with the external CPU for the  data storage.

A trigger algorithm designed to look for clusters of signals over threshold was developed and worked on a PDM basis \citep{SPB2Trigger}. The trigger may come either from the photon counting, or from the charge integration data. A trigger and timing board, called the Clock Board, implements the reference clock distribution and handle the generation of the global trigger. This board collects and elaborates the trigger generated by each PDM and generates a global trigger that starts an event acquisition on all PDMs. Moreover it handles the busy signals, disabling the trigger generation during the data readout, and implements the live and dead time counters. The clock board also maintained the time synchronization of the entire telescope and performed the timing and position tagging of each triggered event using data from two differential GPS receivers. Upon reception of a global trigger signal, the Zynq board stores 128 GTUs, 60 GTUs before and 67 GTUs after the trigger, from both the photon counting and the charge integrated data.

Once the data were recovered off of the FPGA boards, it was then packaged and transmitted to ground by one of two redundant CPUs. Several parallel telemetry streams were flown. In addition to the Tracking Data Relay Satellite System (TDRSS) that has been used on previous ULDB flights, the 2023 campaigns were the first to fly a Starlink connection. Utilizing low-Earth orbit satellites, Starlink provides a connection to the payload with significantly lower latency than geostationary options such as TDRSS. Further, the total bandwidth available through Starlink was roughly 100$\times$ that available through TDRSS. Additionally, two more streams utilizing Iridium satellites were available for commanding and limited data transfer. 

\subsection{Optics}

The three PDMs sat at the focus of a 1~m diameter telescope with a modified Schmidt design. Six rectangular mirror segments made up the reflective area of the telescope. The telescope had a wider FoV than the PDMs are sensitive to. The six mirror segments had many degrees of freedom and could be repositioned relative to one another and the mechanical structure of the telescope without machining. The mechanical structure of the telescope was a c-frame made out of aluminum, with the mirror mounting mechanisms connected to the back and a black anodized aluminum camera shelf in the middle. An aspheric corrector plate sat at the entrance of the telescope to correct off-axis aspheric aberrations. A field-flattener which mapped the curved focal surface of the spherical telescope to the flat surface of the MAPMTs followed by a BG-3 filter sat at the focus of the telescope directly in front of the PDMs. The telescope was surrounded by a light-tight box made of thin foam coated with black paint to minimize reflections of scattered light. 

\subsection{Health LED}

A useful tool for understanding the performance of the instrument are the health light emitting diodes (HLEDs).
Each of the HLEDs was flashed every 32 seconds, with each flash consisting of a 2.5 $\mu$s pulse and a 5 $\mu$s pulse, separated by 15 $\mu$s. 
The two HLEDs are offset from one another by 16 seconds, and are programmed to pulse with different intensities.
The frequency of the HLEDs are chosen so that roughly every other file produced, groupings of nine consecutive triggers, should contain an HLED event. 
By having a total of four different intensities for the HLED, a larger portion of the dynamic range of the instrument can be monitored.
Data from the HLEDs are recorded by utilizing the internal triggering system, with no physical link between the HLED systems and the PDMs. 

% \subsection{Telemetry and Power}

% 

% The power system for EUSO-SPB2 telescopes and associated ancillary devices used 6$\times$24~Amp-hour lithium-ion batteries powered by 15$\times$100~Watt solar panels with two custom designed charge controllers as the link between the two. Since the stated mission parameters allowed for flight at latitudes anywhere between $-5^{\text{o}}$ to $-90^{\text{o}}$ the power system was designed with enough overhead to allow for operation at lower latitudes where the nights could become much longer than the days.  
\section{Detector Characterization}\label{secs:characterization}

\subsection{Laboratory Measurements}

Prior to telescope integration, the pieces of the detector underwent extensive testing and characterization in laboratory settings. The optics were tested using a steady state UV light source uniformly illuminating the aperture of the telescope. This uniform illumination was made possible by the use of a 1~m diameter parabolic mirror with a diverging LED ($\lambda_{\text{LED}}=385$~nm) placed at its focus. Measurements using a photodiode of the incident light intensity in front of the telescope's aperture were compared with measurements made at the focus of the telescope to determine that the overall throughput of the system was $0.45\pm0.01$. These measurements also showed that 95\% of the light was confined within a radius of $1.75\pm0.05$~mm. Due to time constraints, this measurement was not repeated for directions other than along the optical axis. 

\begin{figure}[h!]\begin{center}
	\includegraphics[width=\textwidth]{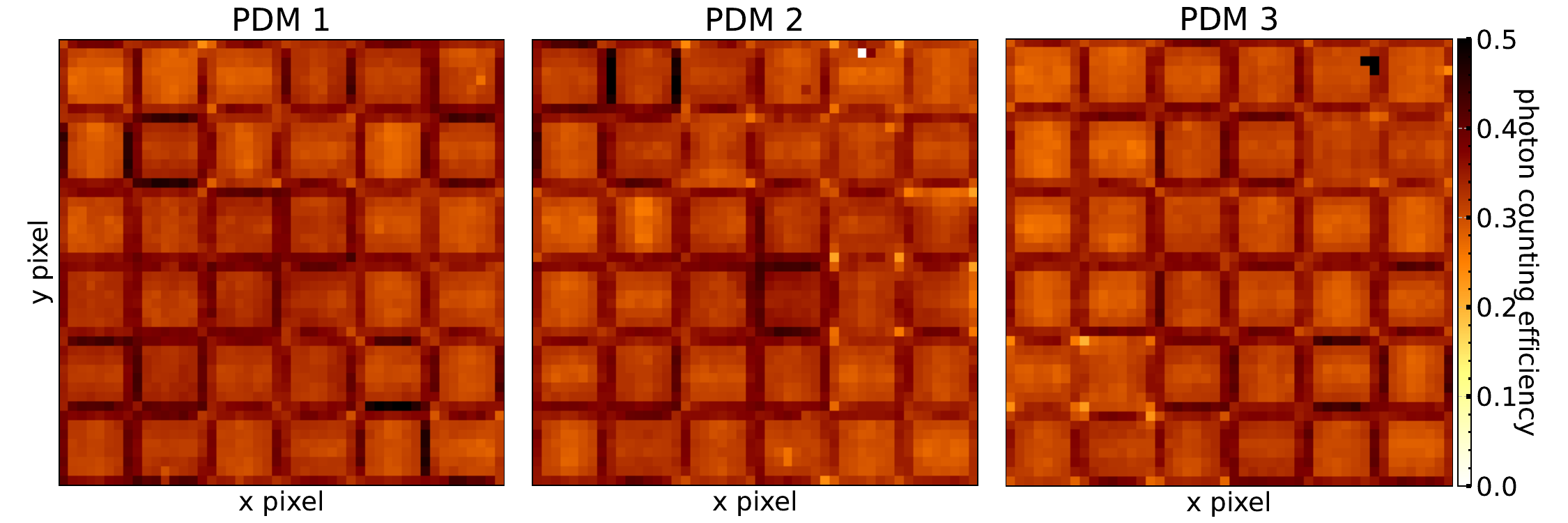}
	\caption{Photon counting efficiency of all pixels under uniform illumination of the entire focal surface, assuming the nominal pixel size. The apparent checkerboard pattern is the result of pixels on the border of MAPMTs being slightly larger and recording more light. The cluster of three pixels in the top right of PDM 3 is the result of three pixels being wired together due to a manufacturing defect.}\label{figs:pix_eff}
\end{center}\end{figure}

The focal surface was also extensively calibrated in the laboratory prior to integration with the telescope optics. Inside of a dark box, a calibrated LED ($\lambda_{\text{LED}}=405$~nm) shined through a pinhole on individual pixels allowed for precise measurements on a pixel-by-pixel basis of important quantities related to the MAPMT's ability to detect EAS. These include the gain, the double-pulse resolution, and the photon-counting efficiency. Further, by moving the LED in submillimeter steps, the shapes of the pixels themselves were able to be mapped. In addition to providing an absolute calibration, these measurements allow for configurable parameters such as the discrimination threshold for counting a ``photo-electron" to be optimized.

The photon counting efficiency under uniform illumination of all 6,912 pixels is shown in Figure \ref{figs:pix_eff}. These efficiencies are calculated using the nominal pixel size provided by the manufacturer. The apparent checkerboard pattern is the result of pixels on the border of MAPMTs being slightly larger and recording more light. The mean efficiency for all pixels is $\mu=0.33$ with a standard deviation of $\sigma=0.03$.

In addition to the laboratory calibrations, the electronics including the data acquisition system and the focal surface were tested in a thermal vacuum chamber.

\subsection{Field Test Measurements}

\begin{figure}[h!]\begin{center}
	\includegraphics[width=\textwidth]{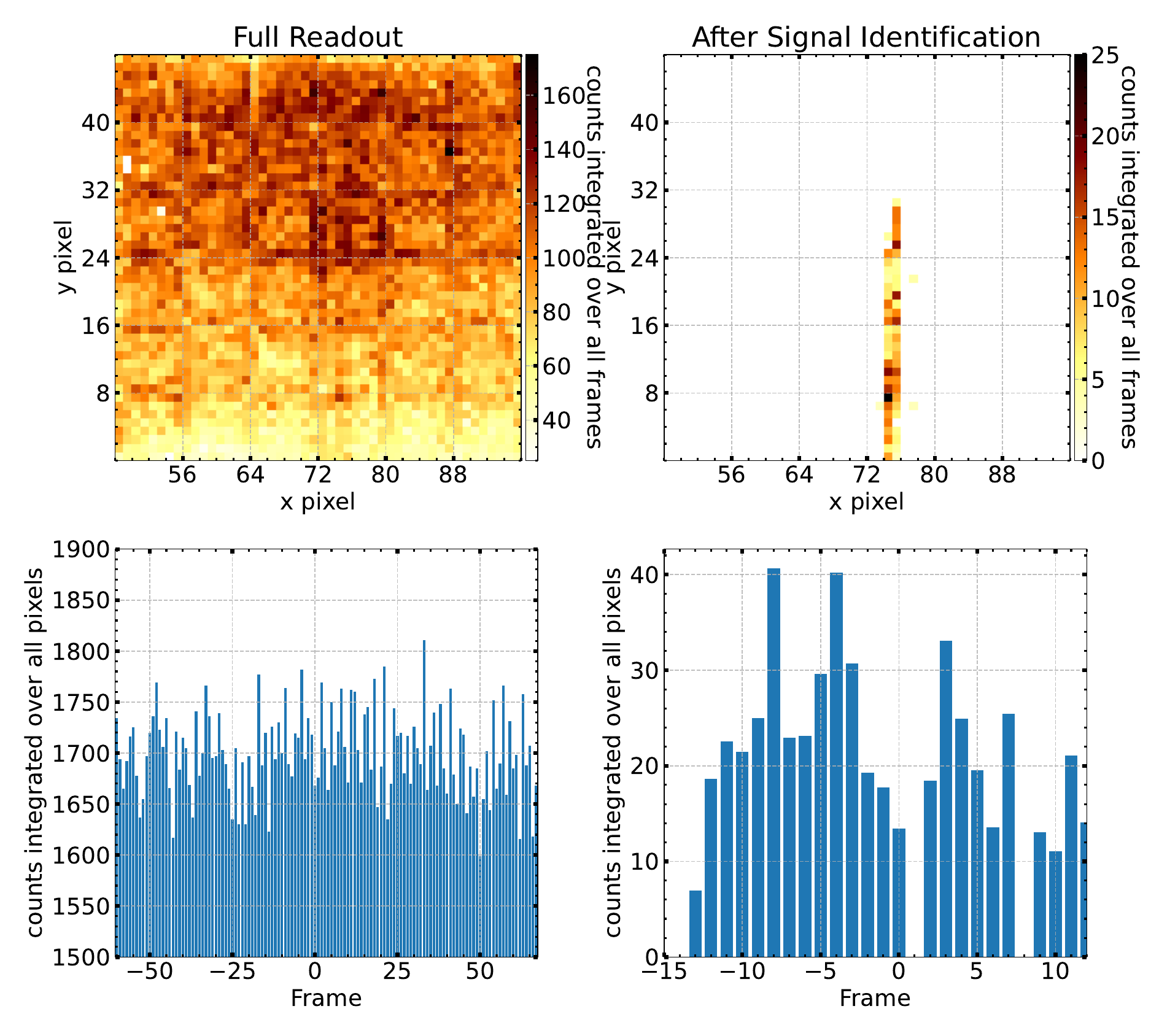}
	\caption{Example of laser recorded during trigger energy threshold scan showing one out of three PDMs. This laser had an energy of 0.89~mJ and an elevation of 45$^\circ$  pointed directly away from the detector which was 24~km away. The left hand column shows the entire 134~$\mu$s readout for one event packet, while the right shows the laser signal after a signal identification algorithm showing the relative scale of the signal of interest. The signal created by this laser corresponds to an EeV scale EAS as seen by the FT from a float altitude of 33~km.}
	\label{figs:laser_example}
\end{center}\end{figure}

In August and September of 2022, the assembled telescope was transported to the Telescope Array Black Rock Mesa Fluorescence Detector \citep{TOKUNO201254} for field testing. The telescope was transported on a mobile trailer in a structure that allowed for pointing in any orientation above the horizon. Over the course of a two-week dark period data were collected utilizing calibrated light source, LEDs and lasers, as well as observations of the night sky including stars, meteors and cosmic rays. This observation campaign allowed for an opportunity to identify and address issues with the system. It also provided data essential for quantifying the end-to-end efficiency of the system and understanding its capabilities in detecting UHECRs.

\begin{figure}[h!]\begin{center}
	\includegraphics[width=\textwidth]{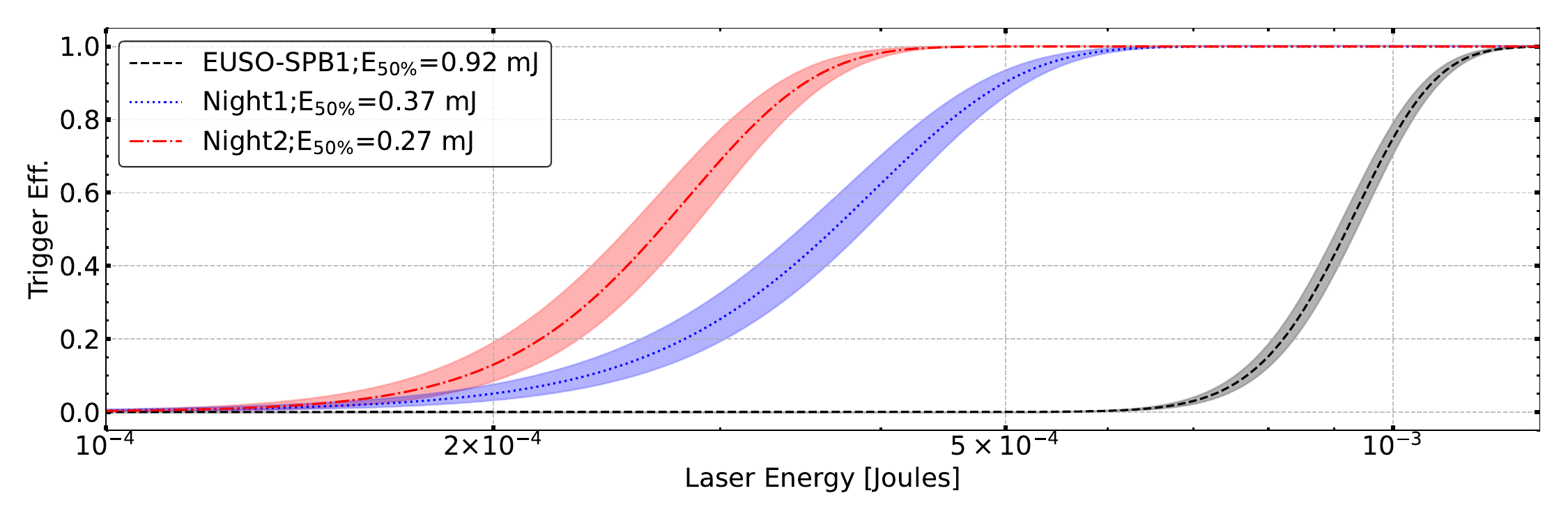}
	\caption{Trigger efficiency as a function of energy. The measurement was repeated on two consecutive nights using the same geometry. The slight difference between the two nights can be attributed to differing background conditions, among other things. The error bands come from the uncertainty in the fit parameters. Both nights show considerable improvement over the previous mission. Data tabulated in Table \ref{tab:lasers}.}
	\label{figs:laser_eff}
\end{center}\end{figure}

A measurement of particular importance is the trigger energy threshold. Due to the nature of the UHECR energy spectrum, a relatively small change in the energy at which the detector triggers can result in a substantial change in the number of events detected. To mimic the optical signature of UHECR induced EAS, a high-power pulsed laser \citep{Hunt:2015a} was used similar to ground based fluorescence detectors \citep{auger_clf,ta_clf}. A frequency tripled NdYAG laser (355~nm) was directed into a steerable periscope style mirror system which is capable of directing the laser in any direction above the horizon. An internal attenuator allows for energies over two orders of magnitude with $E_{\text{Laser}}^{\text{Max}}=90~$mJ. A pick off probe measures a portion of each laser shot via a UV reflecting window to record the energy. The entire laser enclosure is housed in a mobile trailer for easy deployment in the field for testing. 

The goal was to recreate the expected geometry of an extensive air shower observed during the flight when the detector is at 33~km and looking downward. To achieve this, the laser was placed 24~km away and the laser was fired with an elevation of 45$^{\circ}$ directly away from the detector. An example signal from a laser recorded in this geometry is shown in Figure \ref{figs:laser_example}. Lasers were fired in batches of 100 laser shots per energy with energies that were roughly evenly spaced in $\log_{10}(E_{\text{Laser}})$ at a rate of 2~Hz. The telescope was tilted 7$^\circ$ above the horizon so that the entire FoV looked above the ground. The fraction of laser shots that were recorded was then binned in energy and used to calculate a trigger efficiency as a function of energy. This response was fit to a function of the form \begin{equation}\frac{1}{2}\bigg[1+\text{erf}\big((x-a)/b\big)\bigg]\end{equation} where erf is an error function of the form \begin{equation}\text{erf}(z)=\frac{2}{\sqrt\pi}\int_0^ze^{-t^2}dt\end{equation} and $a$ and $b$ are free parameters. The parameter $a$ then gives the energy at which 50\% of lasers triggered.

\begin{table*}
	\begin{center}	
	
	\begin{tabular}{|c|c|c|c||c|c|c|c|}
		\hline
		\multicolumn{4}{|c||}{Night 1}&\multicolumn{4}{|c|}{Night 2} \\
		 \hline
		Energy [mJ] & Fired & Recorded & Ratio & Energy [mJ] & Fired & Recorded & Ratio\\
		\hline

		0.04	&	238	&	8	&	0.03	&	0.04	&	173	&	2	&	0.01\\
0.11	&	297	&	11	&	0.04	&	0.11	&	321	&	13	&	0.04\\
0.18	&	206	&	9	&	0.04	&	0.18	&	181	&	12	&	0.07\\
0.25	&	162	&	10	&	0.06	&	0.25	&	220	&	71	&	0.32\\
0.32	&	101	&	49	&	0.49	&	0.32	&	52	&	47	&	0.90\\
0.39	&	145	&	106	&	0.73	&	0.39	&	90	&	65	&	0.72\\
0.46	&	69	&	53	&	0.77	&	0.46	&	119	&	99	&	0.83\\
0.54	&	95	&	73	&	0.77	&	0.54	&	78	&	64	&	0.82\\
0.61	&	65	&	57	&	0.88	&	0.61	&	69	&	65	&	0.94\\
0.68	&	95	&	83	&	0.87	&	0.68	&	77	&	55	&	0.71\\
0.75	&	80	&	74	&	0.93	&	0.75	&	83	&	74	&	0.89\\
0.82	&	84	&	77	&	0.92	&	0.82	&	70	&	63	&	0.90\\
0.89	&	55	&	52	&	0.95	&	0.89	&	76	&	71	&	0.93\\
0.96	&	80	&	69	&	0.86	&	0.96	&	68	&	68	&	1.00\\
	\hline
	\end{tabular}
	\caption{Number of laser shots recorded and fired during the two nights of trigger energy threshold scans. Laser energy bins in this table ar 70~$\upmu$J wide and linearly spaced. The column labeled Ratio is the trigger efficiency. \label{tab:lasers}}
	\end{center}
\end{table*}

The measurement of the trigger energy threshold was repeated on two separate nights and the results are shown in Figure \ref{figs:laser_eff} and listed in Table \ref{tab:lasers}. In both cases, the threshold was significantly lower than measured for EUSO-SPB1 in field tests during 2016 \citep{Adams_2021}. This was expected due to the higher optical throughput and more sensitive focal surface. There was, however, a significant discrepancy between the theshold energy measured on the two nights with $E^{50\%}_{\text{Night 1}}=0.37\pm0.02$~mJ and $E^{50\%}_{\text{Night 2}}=0.27\pm0.02$~mJ. One contributing factor to this was slightly differing geometries between the two configurations. If the laser track was more centered within a column of pixels then the maximum signal to noise ratio in a pixel would be higher for a given laser energy. This could easily be the result of the azimuth of the telescope being $\approx0.1^\circ$ different between the two nights. This level of discrepancy is not surprising given that the azimuthal pointing of the telescope was not mechanically indexed. Another factor was the differing background conditions, including the intensity of the airglow, between the two energy scans. 

One feature of the data recorded during the trigger energy threshold scan is that even at high energies the trigger efficiency often does not reach 100\%. This is due to other signals in the FoV saturating the DAQ system and causing some lasers to not be recorded. These signals include meteors and airplanes, both of which are very bright compared to EAS. Since neither of these signatures were expected to be observed during flight, the trigger was not designed to reject them. 
\section{Flight overview}\label{secs:flight}
After launch, EUSO-SPB2 had a nominal ascent. 
All subsystems, except for the HVPS, were powered starting before launch and throughout the flight. 
Housekeeping monitoring worked as expected, with measurements of temperatures varying dramatically across the payload. 

% After launch, due to stability issues with the highest bandwidth telemetry connection (Starlink), the download capacity was limited and latency in the communication with the instrument was increased. This is believed to have been caused by a power cycle to the Starlink disk around sunset. The drop in temperature was too great and the current draw from the dish's onboard heater tripped the limit on the PDU channel powering it. The next morning, several hours after sunrise, Starlink was able to turn back on and remained powered for the remainder of the flight.

The in flight commissioning phase for the telescope started once the balloon reached the nominal altitude. During this step, actions were taken to nominally operate the telescope given the actual environmental conditions in flight. The first hours of operation were dedicated to monitoring of the telescope performance. These operations were delayed because of the unavailability of the Starlink system. This unavailability was due to a power failure, and was recovered during daytime. Once recovered, the Starlink system was stable up to the flight termination. This provided the possibility of modifying operating mode and finding the optimal procedures to operate the telescope. Once it was clear that the balloon had serious problems, we had to abort the commissioning phase. The status of the telescope was frozen to get as much science data as possible, until the flight terminated into the southern Pacific Ocean at 12:54 UTC on May 14$^{\text{th}}$ 2023, 36 hours and 52 minutes after launch.

% Upon initializing the first data acquisition at float altitude, it appeared that not all ECs were able to be powered at the nominal high voltage (HV) of 1013 V.
% Re-initializing the instrument several times led to different ECs reaching nominal voltage, with one third to one half of the focal surface being active at a time. 
% This was indicative of the good health of the ECs themselves, but also of the necessity to modify the powering scheme in flight conditions. 
% A slower (safer) increase of the HV during the powering sequence was adopted the second night, which indeed proved fully satisfactory, with all ECs turned on at the intended potential difference, and with nominal efficiency.
% Data taking continued under these conditions for the remainder of the dark period on May 13$^{\text{th}}$, with plans to diagnose the HVPS issues the following night, using full telemetry.

\begin{figure}[h!]\begin{centering}
	\includegraphics[width=\textwidth]{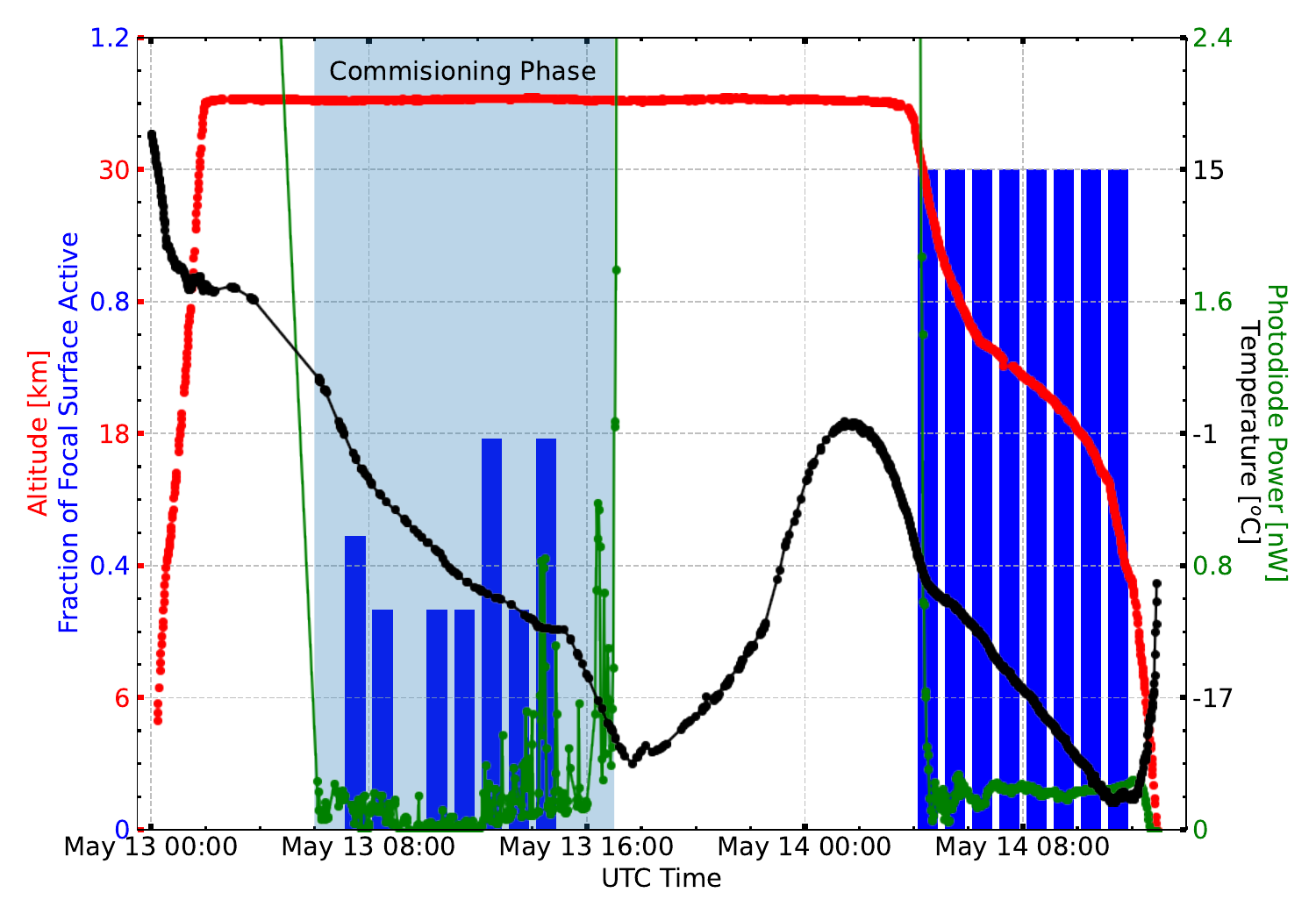}
	\caption{Altitude as a function of time from shortly after launch until flight termination (red),
	fraction of the focal surface active (blue),
	temperature recorded inside of telescope, at the focal surface on the side of the MAPMTs (black),
	light level at focal surface measured by photodiode with a maximum readout of 4.1 nanowatts (green). 
	Note: 0.6 nW is the light level that allows for safe continuous operations of the MAPMTs.}
	\label{figs:prof}
\end{centering}\end{figure}

% After learning of a leak in the balloon on May 14$^\text{th}$, a plan to address the HVPS issue by bringing the photo-cathode to its nominal voltage more slowly was deployed as rapidly as possible. Rather than testing new initialization procedures slowly by sending commands from ground, a new initialization sequence was programmed on ground and pulled directly to the disk of the flight CPU prior to being tested. Normally, this action would not be taken because of the possibility that issues with the new sequence could not be resolved if the Starlink connection was lost once again. However, given the circumstances of an imminently ending flight, the risk was deemed appropriate. 
% By ramping up the HV in 74 steps taking 148 seconds instead of 8 steps taking 16 seconds, all ECs were able to be turned on successfully to their nominal voltage. 
% After performing a threshold scan (often called an s-curve, the instrument operated for the remainder of the flight. 
% With our Starlink connection fully operational, we were able to download the majority of data recorded.
% While the FT was now operating as expected, the balloon rapidly descended, until the flight terminated into the southern Pacific Ocean at 12:54 UTC on May 14$^{\text{th}}$ 2023, 36 hours and 52 minutes after launch.

In total 99,682 triggered events, 134 $\mu$s each, were recorded and downloaded between the two nights. 
All data recorded on May 13$^\text{th}$ were downloaded, and a portion of data recorded on May 14$^\text{th}$ were downloaded. 
The majority of data recorded above 20~km altitude is believed to have been recovered, although due to the circumstances of the flight, this is difficult to verify. 
The time profile of the flight is illustrated in Figure \ref{figs:prof}.

\section{Detector Stability}\label{secs:hled}

\begin{figure}[h!]\begin{center}
	\includegraphics[width=\textwidth]{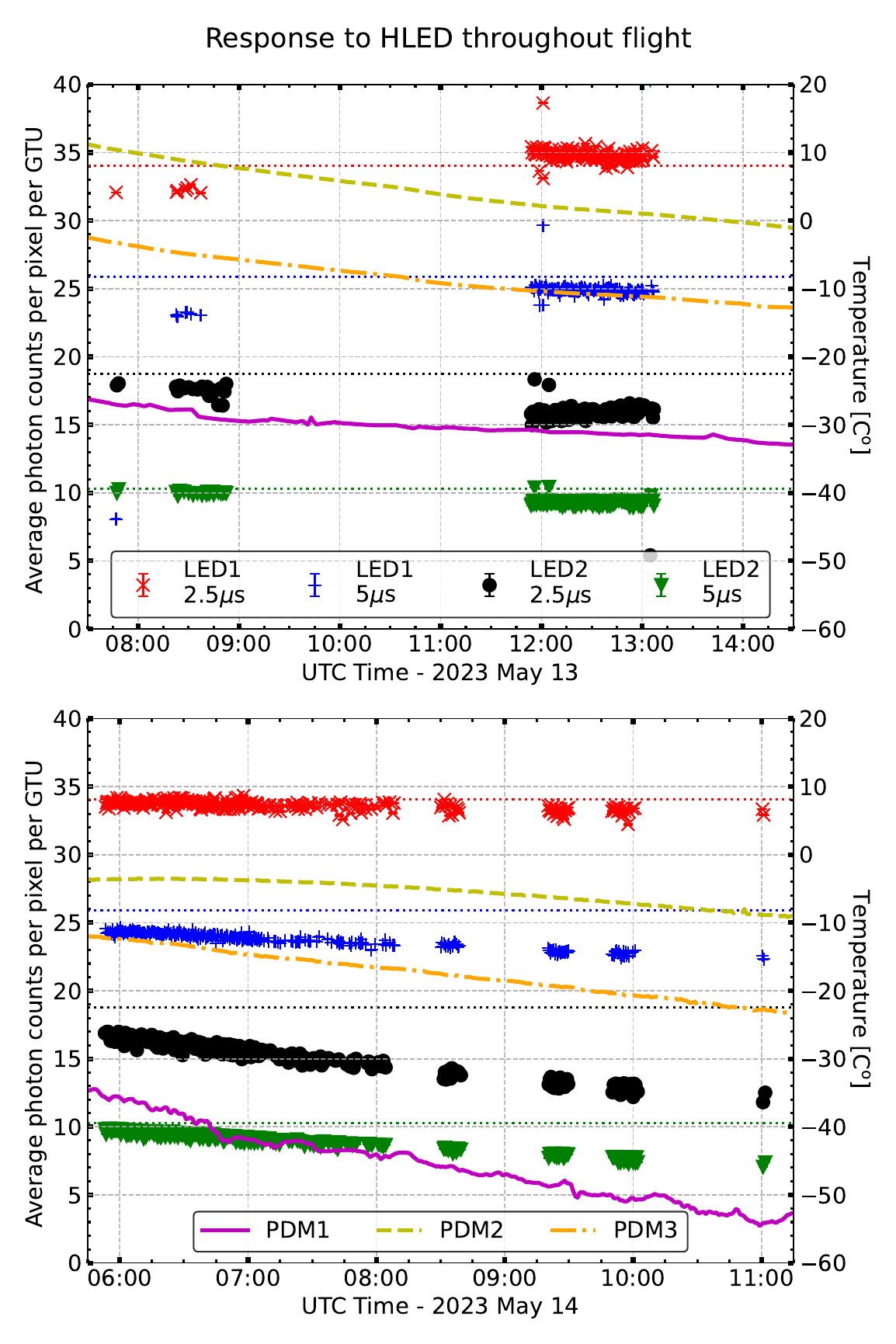}
	\caption{Intensity of HLED flashes over the course of the flight. Average counts per pixel per GTU during each flash. Statistical error bars are plotted but smaller than the marker and not easily visible. Gaps between data on the May 13$^{\text{th}}$ represent time periods where the detector was not active and gaps on May $14^{\text{th}}$ represent data which was recorded but not downloaded. The average intensity measured on ground is plotted for comparison (dotted lines). The temperature is plotted alongside recorded intensity measured at three locations across the focal surface at the MAPMTs (solid and dashed lines).}
	\label{figs:hled}
\end{center}\end{figure}

The response of the system to the LEDs at each intensity over the course of the flight is shown in Figure \ref{figs:hled}.

The lack of HLED data recorded from 09:00 to 11:45 on May 13$^{\text{th}}$ is due to the detector not being active throughout this period due to ongoing diagnosis of the system described in Section \ref{secs:flight}. The missing data between 08:00 and 10:00 on May 14$^{\text{th}}$ is the result of data that were recorded but not downloaded before the termination of the flight. While there are a few outliers on the first night, most likely as a result of the HVPS problems leading to the focal surface being less stable, the majority of flashes are clustered around the intensity of light that was recorded on the ground prior to flight, shown as dashed lines in Figure \ref{figs:hled}. On the second night there are even fewer outliers but there is a consistent downward trend in the recorded light levels. This is believed to be an issue with the HLED and not the detector. The temperature inside of the telescope dropped dramatically during descent and was outside the range in which the HLED systems were designed for and calibrated in. In order to compensate for the output of the HLED being temperature dependant, the circuit which controls the HLED measures the temperature and changes the amount of current supplied such that the light output is kept constant. Since the temperatures experienced were outside the range the device was designed for, the circuit could not provide enough current to mantain a constant light level. The solid lines in Figure \ref{figs:hled} show the measured temperatures at 3 locations across the focal surface. The decrease in the average photon counts per pixel being likely an artifact of HLED rather than the detector's changing response is further confirmed by the relative uniformity of the response across the focal surface.  

\begin{figure}[h!]\begin{center}
	\includegraphics[width=\textwidth]{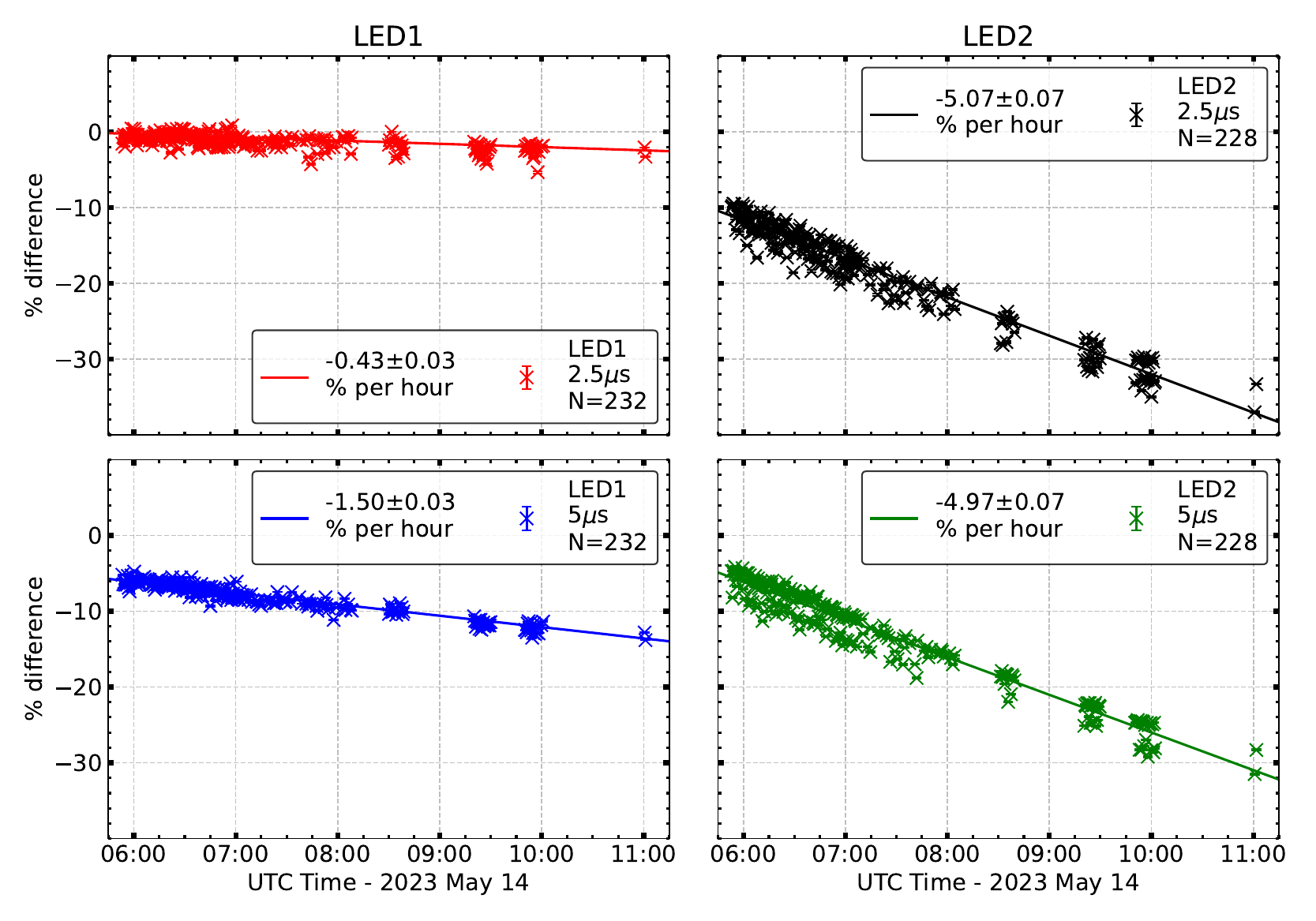}
	\caption{Decrease in intensity of HLED flashes over the course of the second night of the flight measured as a fractional difference compared to measurements on the ground in the weeks before launch. }
	\label{figs:led_time}
\end{center}\end{figure}

The difference in measured intensity compared to the intensity measured on the ground grows with time on the second night, as is visible in Figure \ref{figs:led_time}. The brightest flash, LED1 2.5~$\mu$s shows the smallest decrease. This is due to the flash intensity being near the photon pile-up peak, where the number of photon counts is least sensitive to a change in the incident light. The downward trend for both flashes of LED2 is significantly more pronounced than for LED1, suggesting again that it is in fact an issue with the HLEDs and not the focal surface. In addition to the temperature being lower than expected, there was a significant temperature gradient across the focal surface, as is apparent from the measurements of temperature probes on each PDM, shown in the bottom panel of Figure \ref{figs:hled}. Since the two HLEDs were on different sides of the focal surface, it is possible that LED2 was significantly colder than LED1 leading to the larger decrease in intensity.

\section{Diffuse UV Emissivity}\label{secs:uv}

An important measurement with respect to future missions is the absolute value of the diffuse UV emissivity when looking down at the atmosphere. Measurements have been reported by Mini-EUSO \citep{CASOLINO2023113336}, using similar electronics from the altitude of the international space station (400~km). A one night flight in the year 2000, Nightglow \citep{nightglow}, recorded the diffuse UV background as a function of zenith angle as a precursor for eventual space based measurements of UHECRs. While the asymmetry of their results suggests a misunderstanding of their reported errors, their measurement can still be compared to as a baseline.

\begin{figure}[h!]\begin{center}
	\includegraphics[width=\textwidth]{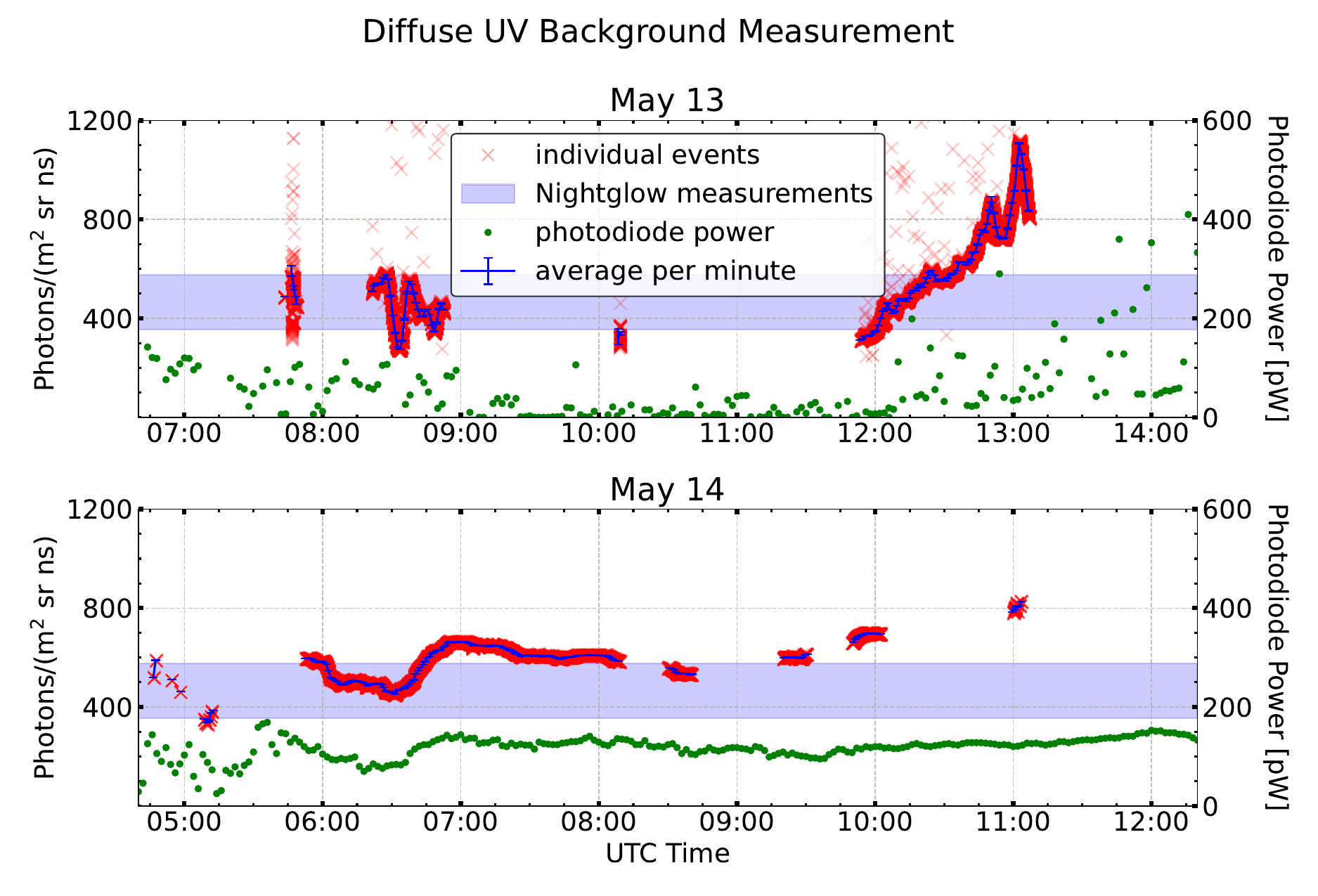}
	\caption{Average light arriving at the aperture of the FT throughout the flight. The red ``x"es show the average light arriving for individual events and the blue error bars show the average of all events in a given minute. Photon count data are converted to photons by using a pixel-by-pixel measurement of the photon counting efficiency of the MAPMTs and measurement of the optical throughput applied to the whole system. Data from an independent photodiode at the focal surface are plotted alongside the FT data (green dots, units on right y-axis).}
	\label{figs:uv}
\end{center}\end{figure}

The average light intensity recorded throughout the flight over all active pixels is shown in Figure \ref{figs:uv}. The red ``x"es show the average for individual events and the blue error bars show the average of all events in a given minute. Events containing HLEDs were not included. There is significantly more dispersion event to event on the first night, an effect that is related to the instrument still being in a commissioning phase as discussed in Section \ref{secs:flight}. The sparsity of the data from the photodiode on the first night is related to issues with telemetry, as this monitoring data was streamed asynchronously from the payload. If a connection was not immediately available the data were not recovered, unlike the data from the main instrument which were downloaded later when telemetry was reestablished. The photon count data recorded by the FT were converted to a flux of physical photons by applying the pixel-by-pixel counting efficiency and a uniform optical efficiency across all pixels, described in Section \ref{secs:characterization}. Since the detector has no way to distinguish different wavelengths of light, and the spectrum of airglow is not precisely known, this calculation does not account for different wavelength photons. The measurement follows the photodiode features closely and is in agreement with the previous measurements reported by Nightglow and the results of EUSO-Balloon \citep{EUSO-Balloon-UV}, despite the arbitrary assumptions of this estimation.

% The measurement follows the photodiode features closely and is in good agreement with the previous measurements reported by Nightglow. Additionally, the measurements are consistent with the results of EUSO-Balloon \citep{EUSO-Balloon-UV}. 

\begin{figure}[h!]\begin{center}
	\includegraphics[width=\textwidth]{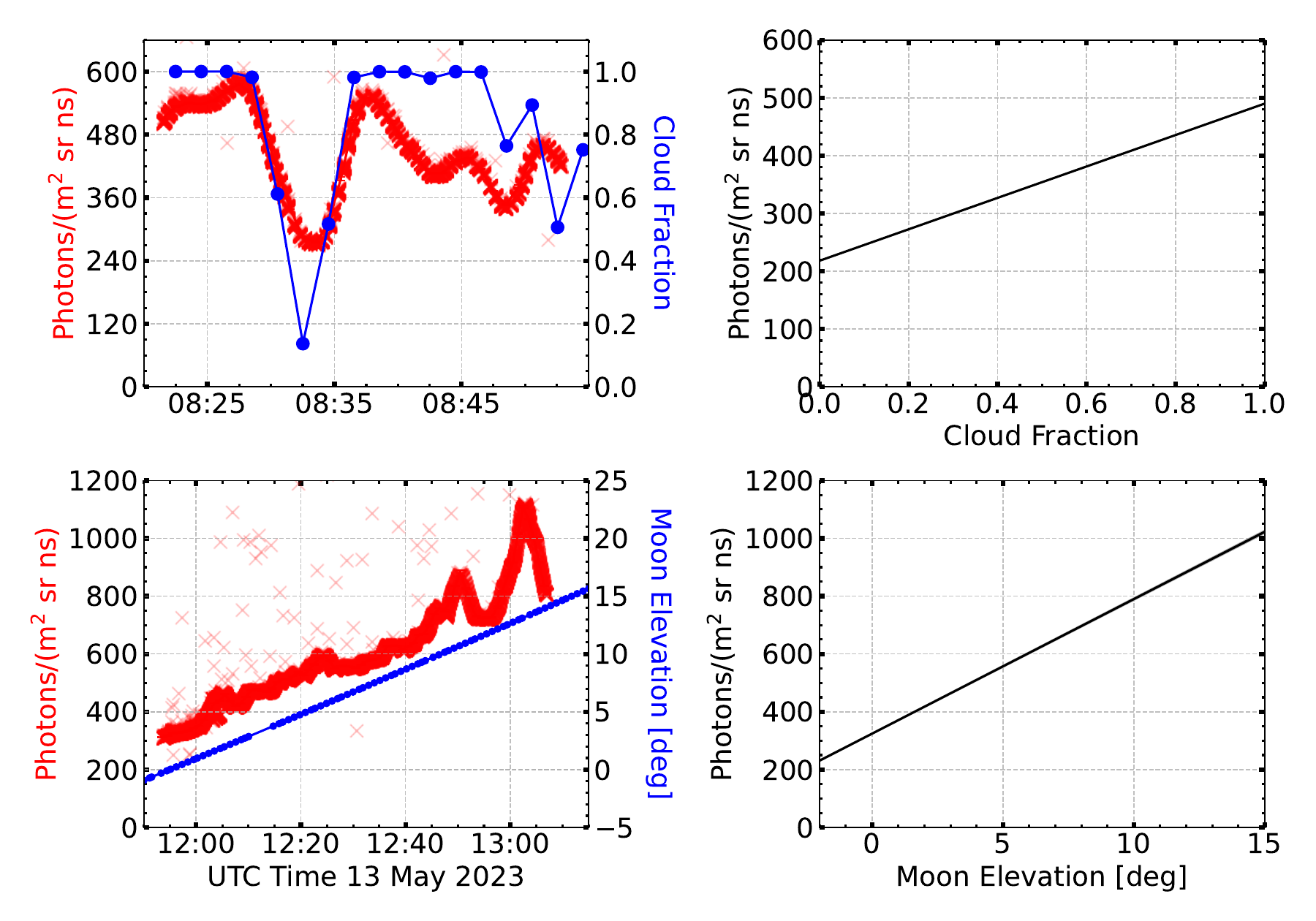}
	\caption{Impact of Moon elevation and cloud fraction on the intensity of UV light observed. Two time periods of interest are shown on the left. The recorded light as a function of cloud fraction (top) and Moon elevation (bottom) are shown on the right.}
	\label{figs:moon}
\end{center}\end{figure}

There is a noticeable increase in the background light intensity after 11:45 UTC on May 13$^{\text{th}}$. This is shown in the lower left panel of Figure \ref{figs:moon}. This increase corresponds to the Moon rising above the horizon. Note: the limb of the Earth is $-5.8^\circ$ relative to horizontal at 33~km altitude, so when the Moon is at any elevation above $-5.8^\circ$ it is ``visible" to the instrument. Once the Moon was $>15^\circ$ above horizontal, the background light was too high to safely operate the MAPMTs and the observations ended. During the period of the Moon rising, the background light recorded increased linearly by $46.5\pm0.2$ photons/(m$^2$~sr~ns) per degree of lunar elevation. The Sun was $>55^\circ$ below the horizon during all observations. 

\begin{figure*}[h!]\begin{center}
	\includegraphics[width=\textwidth]{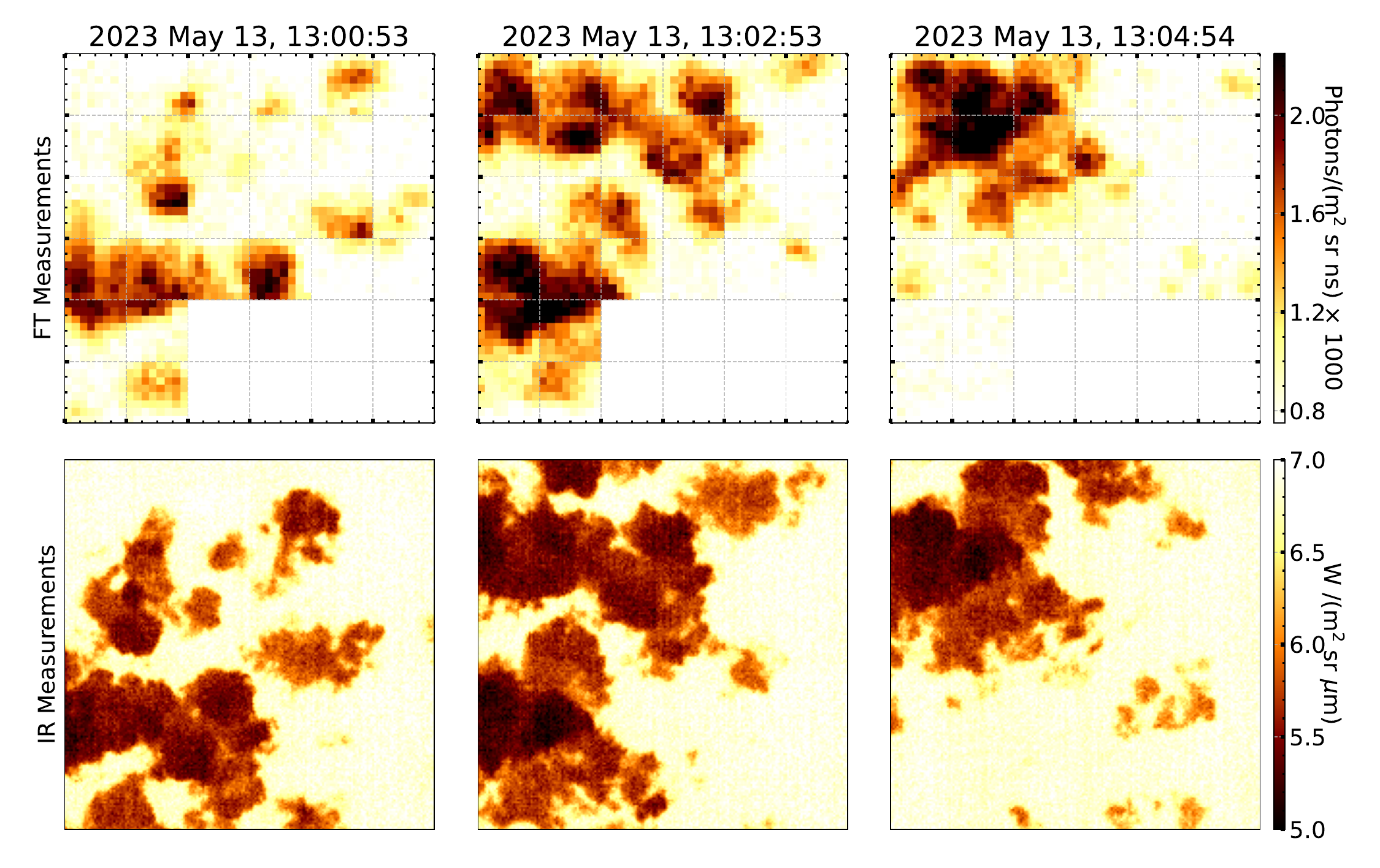}
	\caption{Clouds as seen by the FT and the infrared camera. FT data is averaged over a 134~$\mu$s event readout. IR data is converted to an irradiance by utilizing the known temperature of the ocean based on satellite measurements during the observation window.}
	\label{figs:clouds}
\end{center}\end{figure*}

The fluctuations around the linear rise in background light intensity with Moon elevation angle are attributable to clouds in the FoV. The impact of clouds on the background count rate is more pronounced with the Moon above the horizon as moonlight reflects off the high clouds. The presence of clouds in the FoV can be verified by the independent infrared (IR) camera \citep{Diesing:202345}. Two IR cameras with a $30^\circ\times40^\circ$ FoV (larger than the FT in one direction and roughly equal in the other) image the atmosphere every two minutes with an angular resolution of $\theta=0.08^\circ$ to monitor cloud coverage. By utilizing two different filters, the difference in emissivity can be used to quantify parameters of the cloud including optical depth and cloud top height. In Figure \ref{figs:moon}, cloud fraction is defined as the fraction of pixels with the same FoV as the PDM which have a measured irradiance $>3\sigma$ below the irradiance of the ocean. These measurements are important for constraining the geometric aperture of the instrument. The FT samples at a much faster rate than the IR cameras. Averaging over a 128 GTU event packet allows the structure of clouds to become visible. With the trigger rate around 3~Hz, the clouds are imaged by the FT at a rate much higher than they move. Consecutive events can be used to see the clouds smoothly moving across the FoV, consistent with the slower observations of the IR camera. Comparisons of the FT and IR data in a 4 minute time window are shown in Figure \ref{figs:clouds}.

\section{Exposure Estimation}\label{secs:exposure}

To estimate the exposure of the instrument, extensive simulations are used. These simulations are carried out using the EUSO-\Offline framework \citep{euso-offline}. EAS simulated in CONEX \citep{conex}, using the EPOS LHC hadronic interaction model \citep{epos-lhc}, are the precursor for the simulation. The profile of light, from fluorescence and Cherenkov emission, is calculated and propagated through the atmosphere accounting for Mie and Rayleigh scattering. The photons that reach the telescope's aperture are then projected onto a simulated focal surface using parameters based on measurements described in Section \ref{secs:characterization}, followed by a detailed simulation of the electronics including the trigger algorithm. 

The effective aperture of the FT is less trivially calculated than for ground based arrays. For surface detector arrays, the area where a shower can land and be detected is obviously defined. For an aerial detector looking down at the atmosphere, there are more possible geometries of observable showers. This includes showers which cross sea-level far away from the detector. One method to account for all of the possible geometries of observable showers is to simulate many showers sampling an oversized area beneath the detector and convert the fraction of events which trigger into a triggered aperture by 
\begin{equation}\label{eqs:aperture}
	T=\bigg(\sum_{x\in S}\frac{N^{\text{Triggered}}_x}{N^{\text{Thrown}}_x} \epsilon_x \bigg)A \int_0^{2\pi}d\phi \int_0^{\theta_{\text{max}}}f(\theta)d\theta ,
\end{equation}
where $A$ is the area showers are thrown over and $f(\theta)$ is the distribution of thrown showers. In our case $f(\theta)=\sin\theta\cos\theta$ and $\theta_{\text{max}}=80^\circ$. The quantity $\epsilon_x$ is the energy-dependent UHECR composition fraction based on measurements of the Pierre Auger Observatory \citep{Mayotte:2023Nc}, $x$ is the mass of the primary and $S$ is the set of primary masses simulated $\{\ce{^1H},\ce{^4He},\ce{^{16}O},\ce{^{56}Fe}\}$. Throwing showers over a 100~km radius disk, with an area $A=\pi\times10^4$~km$^2$ covers all possible geometries, with $>$75\% of triggered showers landing within 20~km and $>99$\% landing within 80~km of the center. 

\begin{figure}[h!]\begin{centering}
	\includegraphics[width=\textwidth]{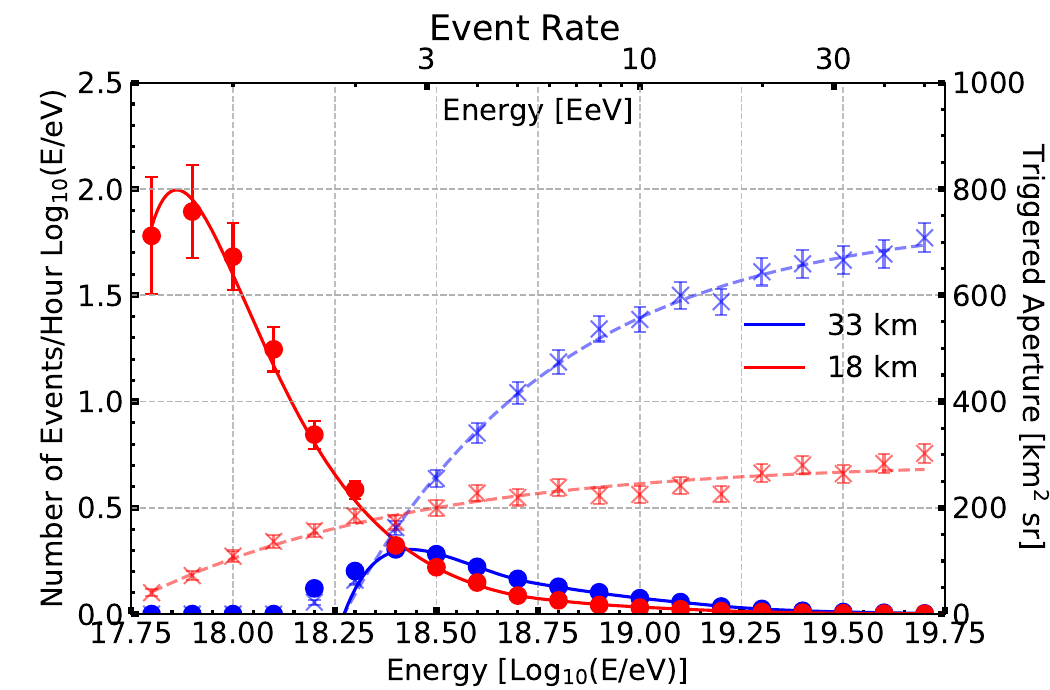}
	\caption{Simulated triggered aperture (dashed lines and ``x''es) and expected event rate (solid lines and circles) at two altitudes 33~km (blue) and 18~km (red). Each curve represents 2 million total simulated EAS.}
	\label{figs:event_rate}
\end{centering}
\end{figure}

Showers are thrown at discrete energies evenly spaced in $\log_{10}(\text{E}/\text{eV})$ at 20 total energies from $10^{17.8}$~eV to $10^{19.7}$~eV. Twenty-five thousand showers of each mass are thrown for a total of 100,000 showers per energy and 2 million showers total. The triggered aperture at each energy is converted into an expected event rate via measurements of the UHECR energy spectrum reported by the Pierre Auger Observatory \citep{Aab_2020} by 
\begin{equation}\label{eqs:rate}
	R(E_i) = T(E_i) \int_{E_i\times10^{-0.05}}^{E_i\times10^{0.05}} J(E)dE ,
\end{equation}
where $J(E)$ is the differential UHECR energy spectrum and $T(E_i)$ is the triggered aperture given by Equation \ref{eqs:aperture}.

Simulated apertures and event rates are shown in Figure \ref{figs:event_rate}. The altitude of the FT impacts both the energy threshold and the aperture. When the payload is lower in the atmosphere, the detector is closer to the emitted light, most of which is created below 10~km altitude, and therefore more of the isotropically emitted fluorescence light is detected. This lowers the minimum energy of EAS that will trigger the instrument. At the same time, at lower altitudes the detector is sensitive to smaller portions of the atmosphere and therefore the effective aperture of the instrument is decreased. Due to the nature of the energy spectrum with the flux of UHECRs decreasing so rapidly with increasing energy, the total number of expected events increases with the decreasing energy threshold despite the decreasing aperture. 

\begin{figure}[h!]\begin{centering}
	\includegraphics[width=\textwidth]{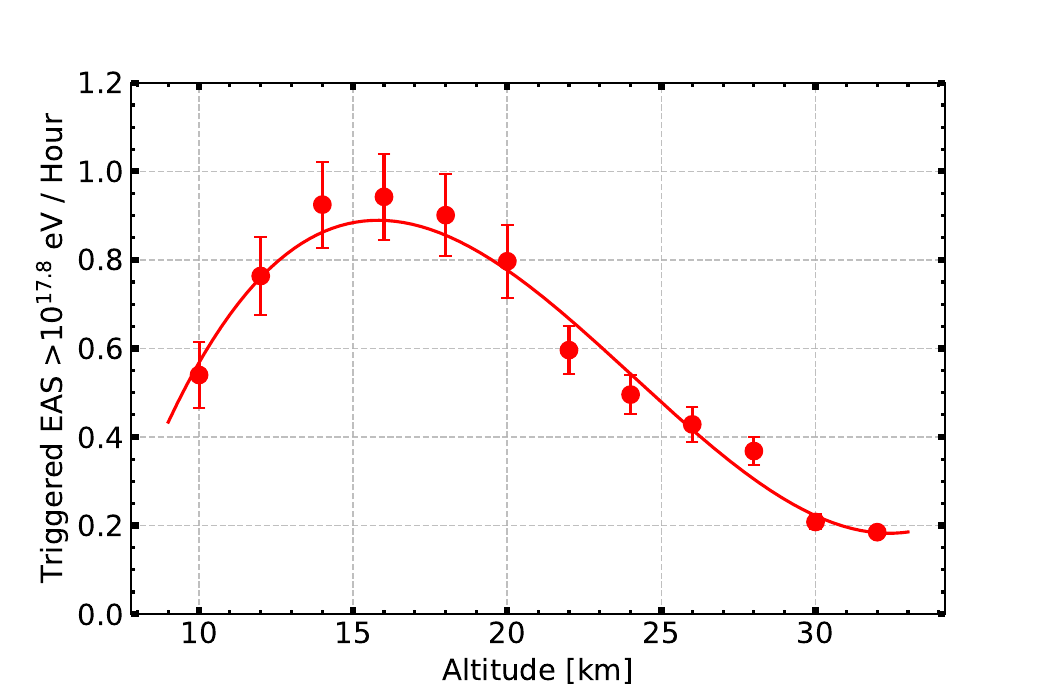}
	\caption{Expected event rate as a function of FT altitude at altitudes from 10-32~km. Each plotted point represents a simulation set of 2~million EAS. The curve is a third degree polynomial used to parameterize the event rate at the different altitudes which the GPS on the balloon recorded data at during the flight. }
	\label{figs:alt_event_rate}
\end{centering}
\end{figure}

Since such a significant fraction of the flight observations were below the planned float altitude of 33~km, the triggered aperture was calculated at several altitudes covering the range of observations. The process described above was repeated at 12 different altitudes from 10-32~km, throwing 2 million showers at each altitude. The altitude dependent event rate is shown in Figure \ref{figs:alt_event_rate}. The expected event rate reaches a maximum around 16~km at around 1 event expected per hour of observation. 

In order to estimate the expected number of events over the entire flight, the HLED can be used as an indicator of when the DAQ system was operating as expected. Since the HLED fired at a regular interval of one flash every 16 seconds, the time in between HLED flashes separated by exactly 16 seconds is when the detector can be considered fully active, including photo-detection, triggering, and crucially, downloading of data to ground. In total, there were 636 periods between consecutive HLED flashes, for a total of 10,716 seconds or 2~hours 58~minutes 36~seconds. Much of this data was recorded below the nominal float altitude of 33~km, requiring the altitude dependent event rate to be used to estimate the total exposure. The total number of expected events over the course of the flight is given by the discrete sum
\begin{equation}\label{eqs:total_exposure}
	\sum_{t_i \in \text{active}} R\big(a(t_i)\big) \times \epsilon(t_i)\times\delta t ,
\end{equation}
where $R\big(a(t_i)\big)$ is the altitude $a(t_i)$ dependent event rate at a time $t_i$, $\epsilon(t_i)$ is the fraction of the focal surface that was active at $t_i$ and $\delta t=16$~seconds is the time period the DAQ was active for based on the measurements of the HLED. Summing over the 636 active periods, the total number of expected events is 1.24. This assumes the nominal background conditions of 1 photon-count/Pixel/GTU, which is consistent with the real background rates observed during the flight. Further, this estimation does not consider the cloud coverage or the impact of the moon since the periods where the moon was above the horizon or clouds were in the FoV represent a relatively small fraction of the overall time period the instrument was actively recording data. The average dead-time per recorded event was 1.1~ms. At the average trigger rate of 7~Hz, this leads to a dead-time fraction of less than 1\% which does not significantly impact the expected total number of events. 
\section{Event Analysis}

\begin{figure}[h!]
    \begin{center}
	    \includegraphics[width=\textwidth]{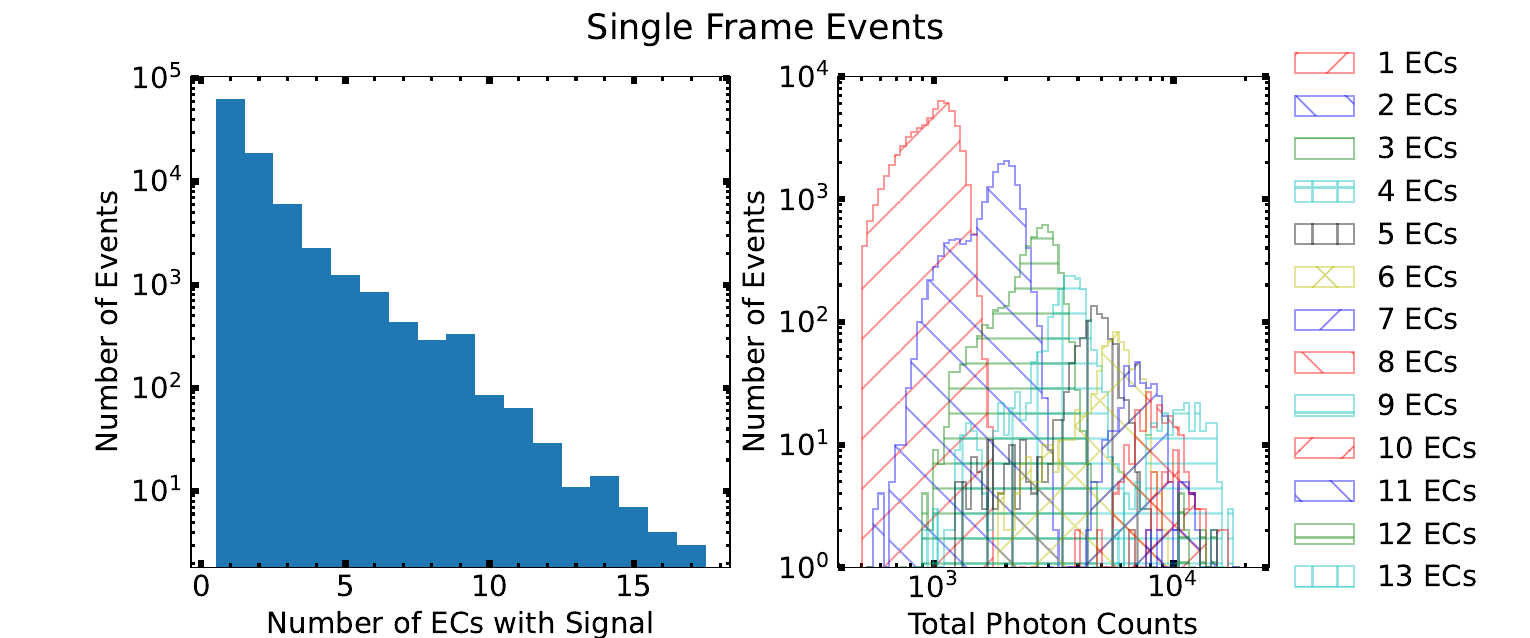}
	    \caption{Distribution of number of ECs with signal for 90,046 single frame events recorded during the flight is shown on the left. The distributions of total signal recorded during each flash for each number of ECs involved in the flash is shown on the right.}
	    \label{figs:ec_flashes}
    \end{center}
\end{figure}

Of the 99,682 triggered and downloaded events, 805 or 0.81\% were events triggered by the HLED system described in Section \ref{secs:hled}. The majority of other events, 90,046 or 90.3\% contained signal that lasted for a single frame. In all cases the signal was localized to a number of elementary cells (ECs, described in Section \ref{secs:desrciption}). That is, if some pixels within an EC had signal, almost all did. Of this population of events, the majority were confined to a single EC, as shown in the left panel of Figure \ref{figs:ec_flashes}. The total signal recorded is normally distributed around a different value depending on the number of ECs with signal. This suggests that rather than recording light from outside of the instrument, these events are detector artifacts that at this time are not fully understood. Very similar events were recorded during the test campaigns on the ground and may be related to electronics or HVPS baseline fluctuations as a response to either direct particle interactions or bright localized light pulses. Since the trigger had several tuneable parameters, these events may have been able to be rejected with some tuning. However, given the circumstances of the flight, there was not enough time to fully optimize the trigger parameters. The limited time afloat was used to record data rather than attempt to reject these signals which were not saturating the readout electronics on the second night.

Many events, 3,500 or 3.5\%, lasted for longer than a single frame and did not increase counts across an entire EC the way the previously described events did. These events appear at first as tracks, however the signal is first visible in every pixel during the same frame and then slowly decreases in all pixels in subsequent frames. These events are believed to be caused by galactic cosmic rays interacting in the photocathode of the MAPMTs and have been observed on previous JEM-EUSO balloon missions.

There were 280 events which lasted for many frames and were very bright. These events contained a signal that moved in the camera with a ring-like structure. Events of this type only ocurred on the first night of the flight, and the ``rings'' of signal were always centered on the edge of an EC which was not at the nominal high voltage. Therefore these events are believed to be a discharge of some type that is related to the high electric field ($\approx$350~kV/m) present between the operating EC and the non-operating neighbor.

\begin{figure}[h!]
    \begin{center}
	    \includegraphics[width=\textwidth]{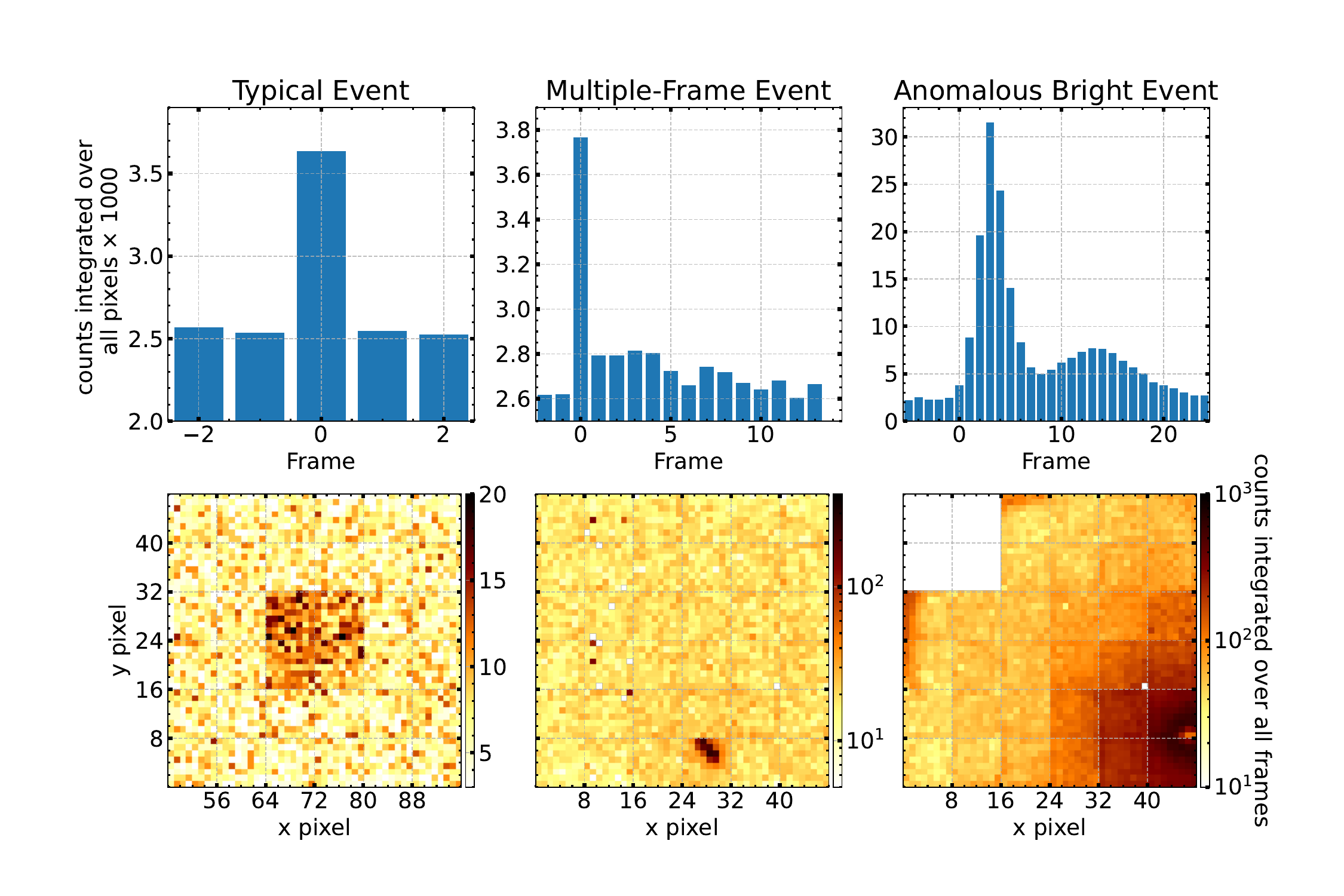}
	    \caption{Three examples of the typical types of events recorded.}
	    \label{figs:example_events}
    \end{center}
\end{figure}

There were 5,051 events or 5.0\% of the total recorded events which did not have any obvious instantaneous signal in them. Rather, the normal fluctuations of the background were enough to satisfy the trigger condition and cause the detector to be readout. This behavior was expected, and after optimization of the trigger parameters would be the expected cause of most triggers. Examples of the classes of events described are shown in Figure \ref{figs:example_events}.

Several independent searches have looked for EAS signatures in the data recorded from the flight. These include the track finding algorithms of EUSO-\OfflineT, a convolution neural network based on simulations described in Section \ref{secs:exposure} and independently designed algorithms. The basis of these algorithms is to search for clusters of signal over background which appear in adjacent pixels moving at a speed which is consistent with an EAS propagating at the speed of light below the detector. Unfortunately no candidate events have been found, which is consistent with the expected number of observations.

\section{Conclusions and Outlook}

After only 36 hours and 52 minutes after launch, EUSO-SPB2 regrettably concluded flight prematurely due to a failure of the balloon. Despite this, the flight represents several technical advancements towards an eventual space-based observatory such as POEMMA. One hundred and eight MAPMTs were triggered and readout in parallel. Their response was verified by the HLED system and observations of clouds in the FoV verified by the independent IR camera. Further, the diffuse UV emissivity of the atmosphere when looking down from above was measured with high precision and the impact of cloud coverage and lunar elevation on the background light level have been quantified. The limited flight duration leads to an expected number of UHECR observations near one, with no candidates identified. The majority of events lasted for a single frame and are most likely detector artifacts which were not adequately rejected by the trigger algorithm due to a shortened commissioning phase. These measurements were recovered primarily thanks to a next-generation low-Earth-orbit satellite internet system with the capability of transforming balloon-based science. In addition to these measurements, valuable technical experience has been gained with regards to thermal management and mission operations, which will be impactful for future missions. Building on the experience gained from the 2023 flight, plans are underway for a successor mission POEMMA Balloon with Radio (PBR) with a flight as early as 2027. 
\section*{Acknowledgements}
This work was partially supported by Basic Science Interdisciplinary Research Projects of RIKEN and JSPS KAKENHI Grant (22340063, 23340081, and 24244042), by the Italian Ministry of Foreign Affairs and International Cooperation, by the Italian Space Agency through the ASI INFN agreement EUSO-SPB2 n. 2021-8-HH.0 and its amendments and through the ASI-INAF agreement n.2017-14-H.O, by NASA award 11-APRA-21730058, 16-APROBES16-0023, 17-APRA17-0066, NNX17AJ82G, NNX13AH54G, 
\hspace{10cm}80NSSC18K0246, 80NSSC18K0473, 80NSSC19K0626, and 80NSSC18K0464 in the USA, by the French space agency CNES, by Slovak Academy of Sciences MVTS JEM–EUSO, by National Science Centre in Poland grant 2020/37/B/ST9/01821, as well as VEGA grant agency project 2/0132/17, and by State Space Corporation ROSCOSMOS and the Interdisciplinary Scientific and Educational School of Moscow University ``Fundamental and Applied Space Research"
This research used resources of the US National Energy Research Scientific Computing Center (NERSC), the DOE Science User Facility operated under Contract No. DE-AC02-05CH11231. 
We acknowledge the contributions of the JEM-EUSO collaboration to the EUSO-SPB2 project.
We thank the Telescope Array collaboration for graciously allowing us to use their facilities for field tests. 

\bibliographystyle{elsarticle-num}
\bibliography{refs}

\begin{thebibliography}{10}
\expandafter\ifx\csname url\endcsname\relax
  \def\url#1{\texttt{#1}}\fi
\expandafter\ifx\csname urlprefix\endcsname\relax\def\urlprefix{URL }\fi
\expandafter\ifx\csname href\endcsname\relax
  \def\href#1#2{#2} \def\path#1{#1}\fi

\bibitem{Aab_2020}
A.~Aab, et~al., Measurement of the cosmic-ray energy spectrum above
  $2.5$x$10^{18}$ using the pierre auger observatory, Physical Review D 102~(6)
  (2020) 062005.
\newblock \href {http://arxiv.org/abs/2008.06486} {\path{arXiv:2008.06486}},
  \href {https://doi.org/10.1103/physrevd.102.062005}
  {\path{doi:10.1103/physrevd.102.062005}}.

\bibitem{auger_exposure}
J.~Abraham, et~al., Trigger and aperture of the surface detector array of the
  pierre auger observatory, Nuclear Instruments and Methods in Physics Research
  Section A: Accelerators, Spectrometers, Detectors and Associated Equipment
  613~(1) (2010) 29–39.
\newblock \href {http://arxiv.org/abs/1111.6764} {\path{arXiv:1111.6764}},
  \href {https://doi.org/10.1016/j.nima.2009.11.018}
  {\path{doi:10.1016/j.nima.2009.11.018}}.

\bibitem{Coleman_2023}
A.~Coleman, et~al., Ultra high energy cosmic rays the intersection of the
  cosmic and energy frontiers, Astroparticle Physics 149 (2023) 102819.
\newblock \href {http://arxiv.org/abs/2205.05845} {\path{arXiv:2205.05845}},
  \href {https://doi.org/10.1016/j.astropartphys.2023.102819}
  {\path{doi:10.1016/j.astropartphys.2023.102819}}.

\bibitem{grand}
J.~Alvarez-Muniz, et~al., The giant radio array for neutrino detection (grand):
  Science and design, Science China Physics, Mechanics; Astronomy 63~(1)
  (2019).
\newblock \href {http://arxiv.org/abs/1810.09994} {\path{arXiv:1810.09994}},
  \href {https://doi.org/10.1007/s11433-018-9385-7}
  {\path{doi:10.1007/s11433-018-9385-7}}.

\bibitem{POEMMA}
A.~Olinto, et~al., The poemma (probe of extreme multi-messenger astrophysics)
  observatory, Journal of Cosmology and Astroparticle Physics 2021~(06) (2021)
  007.
\newblock \href {http://arxiv.org/abs/2012.07945} {\path{arXiv:2012.07945}},
  \href {https://doi.org/10.1088/1475-7516/2021/06/007}
  {\path{doi:10.1088/1475-7516/2021/06/007}}.

\bibitem{Settimo_2016}
M.~Settimo, for~the Pierre Auger~Collaboration, Measurement of the mass
  composition of ultra-high energy cosmic rays with the pierre auger
  observatory, Journal of Physics: Conference Series 718~(5) (2016) 052037.
\newblock \href {https://doi.org/10.1088/1742-6596/718/5/052037}
  {\path{doi:10.1088/1742-6596/718/5/052037}}.

\bibitem{refId0}
{Mayotte, Eric}, {Fitoussi, Thomas}, Update on the indication of a
  mass-dependent anisotropy above 1018.7 ev in the hybrid data of the pierre
  auger observatory, EPJ Web Conf. 283 (2023) 03003.
\newblock \href {http://arxiv.org/abs/2303.16336} {\path{arXiv:2303.16336}},
  \href {https://doi.org/10.1051/epjconf/202328303003}
  {\path{doi:10.1051/epjconf/202328303003}}.

\bibitem{JEM-EUSO}
{Parizot, E}, et~al., The jem-euso program for uhecr studies from space, EPJ
  Web Conf. 283 (2023) 06007.
\newblock \href {https://doi.org/10.1051/epjconf/202328306007}
  {\path{doi:10.1051/epjconf/202328306007}}.

\bibitem{EUSO-Balloon}
J.~Adams, et~al., A review of the euso-balloon pathfinder for the jem-euso
  program, Space Science Reviews 218 (2022) 3.
\newblock \href {https://doi.org/10.1007/s11214-022-00870-x}
  {\path{doi:10.1007/s11214-022-00870-x}}.

\bibitem{Abdellaoui_2018}
G.~Abdellaoui, et~al., First observations of speed of light tracks by a
  fluorescence detector looking down on the atmosphere, Journal of
  Instrumentation 13~(05) (2018) P05023.
\newblock \href {http://arxiv.org/abs/1808.02557} {\path{arXiv:1808.02557}},
  \href {https://doi.org/10.1088/1748-0221/13/05/P05023}
  {\path{doi:10.1088/1748-0221/13/05/P05023}}.

\bibitem{ABDELLAOUI2024102891}
G.~Abdellaoui, et~al., Euso-spb1 mission and science, Astroparticle Physics 154
  (2024) 102891.
\newblock \href {http://arxiv.org/abs/2401.06525} {\path{arXiv:2401.06525}},
  \href {https://doi.org/10.1016/j.astropartphys.2023.102891}
  {\path{doi:10.1016/j.astropartphys.2023.102891}}.

\bibitem{Mini-EUSO}
S.~Bacholle, et~al., Mini-euso mission to study earth uv emissions on board the
  iss, The Astrophysical Journal Supplement Series 253~(2) (2021) 36.
\newblock \href {http://arxiv.org/abs/2010.01937} {\path{arXiv:2010.01937}},
  \href {https://doi.org/10.3847/1538-4365/abd93d}
  {\path{doi:10.3847/1538-4365/abd93d}}.

\bibitem{mini-euso-meteors}
D.~{Barghini}, et~al., Observation of meteors from space with the mini-euso
  detector on board the international space station, Astronomy and Astrophysics
  in pres. (2024).
\newblock \href {https://doi.org/10.1051/0004-6361/202449236}
  {\path{doi:10.1051/0004-6361/202449236}}.

\bibitem{ABDELLAOUI201898}
G.~Abdellaoui, et~al., Euso-ta - first results from a ground-based euso
  telescope, Astroparticle Physics 102 (2018) 98--111.
\newblock \href {https://doi.org/10.1016/j.astropartphys.2018.05.007}
  {\path{doi:10.1016/j.astropartphys.2018.05.007}}.

\bibitem{Gazda:2023cj}
E.~Gazda, {The EUSO-SPB2 Cherenkov Telescope - performance and preliminary
  results}, PoS ICRC2023 (2023) 1029.
\newblock \href {http://arxiv.org/abs/2308.15628} {\path{arXiv:2308.15628}},
  \href {https://doi.org/10.22323/1.444.1029} {\path{doi:10.22323/1.444.1029}}.

\bibitem{PhysRevD.104.063029}
A.~L. Cummings, et~al., Modeling the optical cherenkov signals by cosmic ray
  extensive air showers directly observed from suborbital and orbital
  altitudes, Phys. Rev. D 104 (2021) 063029.
\newblock \href {http://arxiv.org/abs/2105.03255} {\path{arXiv:2105.03255}},
  \href {https://doi.org/10.1103/PhysRevD.104.063029}
  {\path{doi:10.1103/PhysRevD.104.063029}}.

\bibitem{PhysRevD.103.043017}
A.~L. Cummings, et~al., Modeling of the tau and muon neutrino-induced optical
  cherenkov signals from upward-moving extensive air showers, Phys. Rev. D 103
  (2021) 043017.
\newblock \href {http://arxiv.org/abs/2011.09869} {\path{arXiv:2011.09869}},
  \href {https://doi.org/10.1103/PhysRevD.103.043017}
  {\path{doi:10.1103/PhysRevD.103.043017}}.

\bibitem{stratosphere_wind}
D.~J. Karoly, D.~G. Vincent, Meteorology of the Southern Hemisphere, Springer,
  1998.
\newblock \href {https://doi.org/10.1007/978-1-935704-10-2}
  {\path{doi:10.1007/978-1-935704-10-2}}.

\bibitem{romualdez2018overview}
L.~J. Romualdez, et~al., Overview, design, and flight results from superbit: a
  high-resolution, wide-field, visible-to-near-uv balloon-borne astronomical
  telescope, in: Ground-based and Airborne Instrumentation for Astronomy VII,
  Vol. 10702, 2018, p. 107020R.
\newblock \href {http://arxiv.org/abs/1807.02887} {\path{arXiv:1807.02887}},
  \href {https://doi.org/10.1117/12.2307754} {\path{doi:10.1117/12.2307754}}.

\bibitem{KIICRCPaper}
H.~Miyamoto, et~al., {Tests and characterisation of the KI trigger: a trigger
  system for fast events on EUSO-SPB2 Fluorescence Telescope}, PoS ICRC2023
  (2023) 430.
\newblock \href {http://arxiv.org/abs/2310.07388} {\path{arXiv:2310.07388}},
  \href {https://doi.org/10.22323/1.444.0430} {\path{doi:10.22323/1.444.0430}}.

\bibitem{Mini-EUSO_DAQ}
A.~Belov, et~al., {The integration and testing of the Mini-EUSO multi-level
  trigger system}, Advances in Space Research 62~(10) (2018) 2966--2976.
\newblock \href {http://arxiv.org/abs/1711.02376} {\path{arXiv:1711.02376}},
  \href {https://doi.org/10.1016/j.asr.2017.10.044}
  {\path{doi:10.1016/j.asr.2017.10.044}}.

\bibitem{SCOTTI201994}
V.~Scotti, et~al., {The Data Processor system of EUSO-SPB1}, Nuclear
  Instruments and Methods in Physics Research Section A: Accelerators,
  Spectrometers, Detectors and Associated Equipment 916 (2019) 94--101.
\newblock \href {https://doi.org/10.1016/j.nima.2018.10.207}
  {\path{doi:10.1016/j.nima.2018.10.207}}.

\bibitem{SPACIROC3}
S.~Blin, et~al., {SPACIROC3: 100 MHz photon counting ASIC for EUSO-SPB},
  Nuclear Instruments and Methods in Physics Research Section A: Accelerators,
  Spectrometers, Detectors and Associated Equipment 912 (2018) 363--367.
\newblock \href {https://doi.org/10.1016/j.nima.2017.12.060}
  {\path{doi:10.1016/j.nima.2017.12.060}}.

\bibitem{SPB2Trigger}
G.~Filippatos, et~al., Development of a cosmic ray oriented trigger for the
  fluorescence telescope on {EUSO}-{SPB}2, Advances in Space Research 70~(9)
  (2021) 2794--2803.
\newblock \href {http://arxiv.org/abs/2201.00794} {\path{arXiv:2201.00794}},
  \href {https://doi.org/10.1016/j.asr.2021.12.028}
  {\path{doi:10.1016/j.asr.2021.12.028}}.

\bibitem{TOKUNO201254}
H.~Tokuno, et~al., New air fluorescence detectors employed in the telescope
  array experiment, Nuclear Instruments and Methods in Physics Research Section
  A: Accelerators, Spectrometers, Detectors and Associated Equipment 676 (2012)
  54--65.
\newblock \href {http://arxiv.org/abs/1201.0002} {\path{arXiv:1201.0002}},
  \href {https://doi.org/10.1016/j.nima.2012.02.044}
  {\path{doi:10.1016/j.nima.2012.02.044}}.

\bibitem{Hunt:2015a}
P.~Hunt, et~al., {The JEM-EUSO global light system laser station prototype},
  PoS ICRC2015 (2016) 626.
\newblock \href {https://doi.org/10.22323/1.236.0626}
  {\path{doi:10.22323/1.236.0626}}.

\bibitem{auger_clf}
B.~Fick, et~al., The central laser facility at the pierre auger observatory,
  Journal of Instrumentation 1~(11) (2006) P11003.
\newblock \href {https://doi.org/10.1088/1748-0221/1/11/P11003}
  {\path{doi:10.1088/1748-0221/1/11/P11003}}.

\bibitem{ta_clf}
Y.~Takahashi, et~al., {Central Laser Facility Analysis at The Telescope Array
  Experiment}, AIP Conference Proceedings 1367~(1) (2011) 157--160.
\newblock \href {https://doi.org/10.1063/1.3628734}
  {\path{doi:10.1063/1.3628734}}.

\bibitem{Adams_2021}
J.~H. Adams, et~al., Extreme universe space observatory on a super pressure
  balloon 1 calibration: from the laboratory to the desert, Experimental
  Astronomy 52~(1-2) (2021) 125--140.
\newblock \href {http://arxiv.org/abs/2011.09617} {\path{arXiv:2011.09617}},
  \href {https://doi.org/10.1007/s10686-020-09689-2}
  {\path{doi:10.1007/s10686-020-09689-2}}.

\bibitem{CASOLINO2023113336}
M.~Casolino, et~al., Observation of night-time emissions of the earth in the
  near uv range from the international space station with the mini-euso
  detector, Remote Sensing of Environment 284 (2023) 113336.
\newblock \href {http://arxiv.org/abs/2212.02353} {\path{arXiv:2212.02353}},
  \href {https://doi.org/10.1016/j.rse.2022.113336}
  {\path{doi:10.1016/j.rse.2022.113336}}.

\bibitem{nightglow}
L.~M. {Barbier}, et~al., {NIGHTGLOW: an instrument to measure the
  Earth{\textquoteright}s nighttime ultraviolet glow{\textemdash}results from
  the first engineering flight}, Astroparticle Physics 22~(5-6) (2005)
  439--449.
\newblock \href {https://doi.org/10.1016/j.astropartphys.2004.10.002}
  {\path{doi:10.1016/j.astropartphys.2004.10.002}}.

\bibitem{EUSO-Balloon-UV}
G.~Abdellaoui, et~al., Ultra-violet imaging of the night-time earth by
  euso-balloon towards space-based ultra-high energy cosmic ray observations,
  Astroparticle Physics 111 (2019) 54--71.
\newblock \href
  {https://doi.org/https://doi.org/10.1016/j.astropartphys.2018.10.008}
  {\path{doi:https://doi.org/10.1016/j.astropartphys.2018.10.008}}.

\bibitem{Diesing:202345}
R.~Diesing, et~al., {Infrared Cloud Monitoring with UCIRC2}, PoS ICRC2023
  (2023) 450.
\newblock \href {http://arxiv.org/abs/2310.08607} {\path{arXiv:2310.08607}},
  \href {https://doi.org/10.22323/1.444.0450} {\path{doi:10.22323/1.444.0450}}.

\bibitem{euso-offline}
S.~Abe, et~al., Euso-offline: A comprehensive simulation and analysis
  framework, Journal of Instrumentation 19~(01) (2024) P01007.
\newblock \href {http://arxiv.org/abs/2309.02577} {\path{arXiv:2309.02577}},
  \href {https://doi.org/10.1088/1748-0221/19/01/P01007}
  {\path{doi:10.1088/1748-0221/19/01/P01007}}.

\bibitem{conex}
T.~Bergmann, et~al., One-dimensional hybrid approach to extensive air shower
  simulation, Astroparticle Physics 26 (2007) 420--432.
\newblock \href {http://arxiv.org/abs/astro-ph/0606564}
  {\path{arXiv:astro-ph/0606564}}, \href
  {https://doi.org/10.1016/j.astropartphys.2006.08.00}
  {\path{doi:10.1016/j.astropartphys.2006.08.00}}.

\bibitem{epos-lhc}
T.~Pierog, et~al., Epos lhc: Test of collective hadronization with data
  measured at the cern large hadron collider, Phys. Rev. C 92 (2015) 034906.
\newblock \href {http://arxiv.org/abs/1306.0121} {\path{arXiv:1306.0121}},
  \href {https://doi.org/10.1103/PhysRevC.92.034906}
  {\path{doi:10.1103/PhysRevC.92.034906}}.

\bibitem{Mayotte:2023Nc}
E.~Mayotte, et~al., {Measurement of the mass composition of ultra-high-energy
  cosmic rays at the Pierre Auger Observatory}, PoS ICRC2023 (2023) 365.
\newblock \href {https://doi.org/10.22323/1.444.0365}
  {\path{doi:10.22323/1.444.0365}}.

\end{thebibliography}

\end{document}